\documentclass[a4paper,11pt]{article}
\usepackage{pos}
\usepackage{hyperref}
\usepackage{gensymb}
\usepackage{enumitem}
\usepackage{tikz}
\usetikzlibrary{mindmap,shadows}

\renewcommand{\logo}{\relax}

\title{Quantum Gravity in 30 Questions}

\author*[a]{R. Loll}
\author[b]{G.\ Fabiano}
\author[b]{D.\ Frattulillo}
\author[c]{F.\ Wagner}

\affiliation[a]{Institute for Mathematics, Astrophysics and Particle Physics, Radboud University \\ 
Heyendaalseweg 135, 6525 AJ Nijmegen, The Netherlands}

\affiliation[b]{Dipartimento di Fisica Ettore Pancini, Universit\`a di Napoli ``Federico II", \\
Complesso Univ.\ Monte S.\ Angelo, 80126 Napoli, Italy}

\affiliation[c]{Institute of Physics, University of Szczecin,\\Wielkopolska 15, 70-451 Szczecin, Poland \\ \vspace{0.4cm}}

\emailAdd{r.loll@science.ru.nl}
\emailAdd{giuseppe.fabiano@unina.it}
\emailAdd{domenico.frattulillo@unina.it}
\emailAdd{fabian.wagner@usz.edu.pl}

\abstract{\vspace{1cm}
{\bf Abstract.} Quantum gravity is {\it the} missing piece in our understanding of the fundamental interactions today. 
Given recent observational breakthroughs in gravity, providing a quantum theory for what lies beyond general relativity is more urgent than ever.
However, the complex history of quantum gravity and the multitude of available approaches can make it difficult to get a grasp of the topic and its main challenges and opportunities.
We provide a guided tour of quantum gravity in the form of 30 questions, aimed at a mixed audience of learners and practitioners.
The issues covered range from basic motivational and background material to a critical assessment of the status quo and future of the subject. 
The emphasis is on structural issues and our current understanding of quantum gravity as a quantum field theory of dynamical geometry beyond perturbation theory.
We highlight the identification of quantum observables and the development of effective numerical tools as critical to future progress. 
}

\FullConference{%
  Corfu Summer Institute 2021 "School and Workshops on Elementary Particle Physics and Gravity"\\
  29 August - 9 October 2021\\
  Corfu, Greece
}

\tableofcontents

\begin{document}
\maketitle

\section*{Introduction}
\addcontentsline{toc}{section}{Introduction}

The search for a quantum theory of gravity has attracted generations of researchers in fundamental physics, and continues to fascinate and challenge us. The lack of an established theory of quantum gravity means that gravity remains the least well understood of the four fundamental interactions. However, there is more reason than ever to believe that this unsatisfactory situation can and will change.  
Interest in gravity and its foundations is booming, and the opening of new observational channels, most importantly, the direct detection of gravitational waves, is giving fresh impetus to the quest for quantum gravity and its possible phenomenological ramifications. 

It is not easy for newcomers and outsiders to understand what quantum gravity is all about. On the one hand, considerable background knowledge (of general relativity, quantum field theory and more) is needed to grasp many of the technical and conceptual difficulties of the subject.
On the other hand, quantum gravity is a {\it theory in progress}, with a large variety of approaches that are based on vastly different ingredients and assumptions. Without direct links to observation or experiment, 
it is not straightforward to distinguish between promising and misguided ideas.
Despite this difficulty and the uncertainties of what lies beyond the realm of perturbative quantum field theory, there is no shortage of strong opinions among practitioners on what the correct route to quantum gravity ought to be \cite{Armas2021}. 
A constructive way to address this situation is the further development of effective nonperturbative computational tools. 
Currently available in only a few approaches, they enable us to transcend prima facie arguments about the suitability of the various
formalisms and instead focus on a comparison of {\it results}, namely, the computational outcomes of concrete, invariantly 
defined observables. This is not just an idle exercise, since being able to compute ``numbers'' is also essential for deriving new 
predictions, whatever your favourite candidate theory of quantum gravity may be. 

These notes are based on four hours of lectures, designed and delivered by one of us (RL) at the Training School on ``Quantum Gravity Phenomenology in the Multi-Messenger Approach'' for master students, PhD students and early-stage postdocs on the beautiful island of Corfu in the autumn of 2021. Their aim was to provide an introduction and overview of quantum gravity to a mixed audience in terms of prior exposure to the subject. Given the constraints on the time available, such an exposition could not comprehensively cover the long history and complex technical nature of quantum gravity, but had to rely on a selection of background material, topics and ideas. To nevertheless achieve a reasonable breadth and keep the listeners engaged, the material was presented in the form of 30 ``bite-sized'' questions and answers on quantum gravity. They ranged from bread-and-butter issues on motivation, classical origins and certain technical concepts to more subtle ones, e.g.\ the difficulties encountered in quantization and the role of reduced models, 
without losing sight of the bigger picture and 
an assessment of the future of the field. Together they constitute a guided tour of the subject, providing an overall context and a starting point for further exploration. The answers to the 30 questions reflect the perspective of the first author, a long-term researcher in nonperturbative quantum gravity, and as such do not claim to be either unbiased or complete.
The scientific views expressed in these notes are hers, while the co-authors assisted in assembling the material and giving feedback.  

In this write-up of the Corfu lectures, we have adhered to the questions format and kept the style and content at a relatively light level, outlining the logic of an argument rather than providing explicit derivations and technical details. 
We hope that an interested novice will find this an inspiring first encounter with the subject, and will be able to dive deeper into the issues by consulting the literature references provided. We have included a fair number of references, but the field is vast and our
selection necessarily eclectic. The bibliography is by no means comprehensive\footnote{An excellent resource that tries to
present quantum gravity in its entirety is the book by Kiefer \cite{Kiefer2012}.}, and we apologize to those whose
contributions are not mentioned. 
We also hope that researchers already active in the field will feel stimulated to critically examine our and their own answers to the 30 
questions posed, and will formulate new and constructive questions that can serve as stepping stones to further progress in quantum gravity.

 \section*{Q1: What is quantum gravity?}
 \label{sq1}
 \addcontentsline{toc}{section}{Q1: What is quantum gravity?}

{\it Quantum Gravity} is the name for the still elusive fundamental quantum theory underlying Einstein's classical field theory of general relativity. As such it should provide the fundamental laws and equations of motion describing the behaviour of spacetime and gravitational interactions across a vast range of length scales, from at least the tiny Planck length 
\begin{equation}
\ell_{\rm P}=\sqrt{\frac{G_{\rm N}\hbar}{c^3} }\approx 1.616\times 10^{-35}m
\label{lpl}
\end{equation} 
to the diameter of the observable universe at about $8.8\times 10^{26} m$. 

Gravity plays different roles, depending on the scale under consideration. Due to the long-range character of the gravitational force and its universal, attractive nature, gravity forms an essential part of the dynamics of our universe at large, as described by standard cosmology. From cosmic to (sub-)millimeter scales, general relativity appears to give an excellent description of gravitational phenomena consistent with all observations to date, from large-scale astrophysics to black-hole detections to solar-system tests \cite{Will2018,Ligo2021,Ishak2019}. Well-known aspects of Newtonian and Einsteinian gravity are being verified by desktop experiments at ever smaller scales, including the effect of {\it mm}-sized ``point''' sources \cite{Westphal2020} and the inverse-square law at $\mu m$-distances \cite{Lee2020}. 

Proceeding to shorter, atomic and subatomic scales, the laws of general relativity should still be adequate. However, gravity seems to be completely insignificant on these scales since physical processes are dominated by the much stronger electromagnetic and nuclear forces of the standard model of particle physics. This includes the smallest distances ($\approx 10^{-19} m$) that can be resolved by today's particle accelerators,
beyond which we enter a rather speculative realm. 
According to effective quantum field theory lore, gravity in the form of general relativity may well coexist with other quantum fields for many of the 16 orders of magnitude in distance scale down to the Planckian regime \cite{Donoghue2012}. Be that as it may, as the Planck length $\ell_{\rm P}$ is approached, the limit of applicability of Einstein's theory will eventually be reached, in the sense that quantum fluctuations of spacetime -- described by a quantum theory of gravity -- must be taken into account in all processes. Although the Planck scale is quantum gravity's primary habitat, this does not mean that quantum-gravitational effects are confined to this scale. 
Indeed, in a universal theory of quantum gravity, one hopes to identify consequences of the quantum nature of gravity on much larger scales, which despite the weakness of the gravitational force are experimentally or observationally accessible (see \cite{DeWitt1957} for an early discussion). With regard to the latter, in order to make best use of the wealth of data from current and future gravitational wave detectors, there is an urgent need for
theorists to complete their theories of quantum gravity, to provide input to the experimentalists' search for what lies beyond general relativity.

\section*{Q2: Why are we interested in quantum gravity?}
\label{sq2}
\addcontentsline{toc}{section}{Q2: Why are we interested in quantum gravity?}

There are many questions about spacetime and gravity that general relativity is unable to answer, the most obvious ones pertaining to extreme distance and energy regimes. The generic occurrence of singularities in solutions to the Einstein equations, like those inside black holes or at the beginning of our universe (see \cite{Senovilla2014} for a recent review), means that the classical theory ``predicts its own downfall", with the expectation that a quantum theory of gravity is needed to consistently describe such regimes and uncover the true (and presumably nonsingular) nature of these classical singularities. More specifically, we want to understand what happens near the singularity at the heart of a black hole, and we would like to trace the history of the very early universe further back in time to understand the possible quantum origins of cosmological structure formation and of spacetime itself. 

Researchers have been speculating for a long time which aspects, if any, of classical spacetime persist at the Planck scale, and therefore which microscopic degrees of freedom are suitable to capture its quantum dynamics. Many nonperturbative approaches to quantum gravity abandon the Lorentzian four-metrics of general relativity, 
since they are not considered well suited to describing a strongly quantum-fluctuating spacetime. Frequently invoked images of the latter are that of a {\it spacetime foam} full of spikes and wormholes, or of a fundamentally discrete structure made of elementary {\it spacetime atoms}, both of them far removed from the smooth metric fields of the classical formulation.

A driving force behind research in quantum gravity is the widely shared conviction among theoretical high-energy physicists that ``there must be something rather than nothing'', namely, a microscopic quantum theory underlying gravity, just like for the other fundamental forces, never mind how much its ingredients and dynamical laws resemble those of the classical gravitational theory, and whether it is part of some grand unified theory of all the interactions. A counterweight to the difficulties that have already been encountered in the search for quantum gravity is the hope and expectation that it may provide answers to some of the big questions about the origin of space, time and the universe, which have the potential to reshape our understanding of the foundations of physics and our place in the world.

\section*{Q3: Why do we not have a quantum theory of gravity yet?}
\label{sq3}
\addcontentsline{toc}{section}{Q3: Why do we not have a quantum theory of gravity yet?}

Starting from general arguments, one underlying reason is the extreme weakness of gravity, as expressed by the smallness of the dimensionful gravitational constant (Newton's constant)
\begin{equation}
G_{\rm N}\approx 6.67\times 10^{-11}\frac{m^3}{\mathit kg\hspace{0.05 cm} s^2}.
\label{newton}
\end{equation}
To illustrate the situation on atomic scales, the gravitational attraction between two protons, which is proportional to $G_{\rm N}$, is a factor $10^{36}$ smaller than the magnitude of the Coulomb force between them! Also, the smallness of the {\it scale of quantum gravity} $\ell_{\rm P}$ is directly related to the weakness of gravity, as can be seen from eq.\ (\ref{lpl}). Practically speaking, almost everywhere in physics one can therefore get away with either ignoring gravity or using general relativity, mostly to describe the dynamics of large astrophysical objects and without any obvious need for a quantum treatment. Gravity's weakness also contributes to the fact that there are currently no measurements crying out for a quantum-gravitational explanation. From a purely utilitarian point of view, one may therefore regard quantum gravity as an esoteric subject of little urgency.  

However, even aficionados of quantum gravity must address the formidable challenge of finding the correct theory without guidance from observation and experiment, something that has played a big role in the development of all other physical theories. The prospects for obtaining data are definitely improving, with impressive advances of multi-messenger astrophysics in observing strong-gravity events \cite{Addazi2021} and multiple efforts to devise tabletop experiments that may probe the quantum nature of gravity \cite{Bose2017,Marletto2017}. How far we are from any breakthrough results is difficult to say; they will need not only more accurate measuring instruments, but a better theoretical grasp of quantum gravity, to enable us to make quantitative predictions of specific quantum signatures. 
A fruitful strategy to bring about the latter is to develop robust and quantitative criteria to test and compare existing candidate theories, 
rather than pile up conjectures about exotic scenarios at the Planck scale. This will be spelled out in more detail in the second half of the lecture notes.   

In terms of more technical issues, progress in quantum gravity has been hampered by the fact that beyond perturbation theory, where gravity turns out to be nonrenormalizable \cite{Goroff1985}, it does not fit the mould of standard relativistic quantum field theory. The key difference in gravity lies in the fact 
that spacetime itself is dynamical, and its local curvature structure -- in a classical context usually described by metric field variables -- is not fixed a priori, but
determined by solving the equations of motion. By contrast, spacetime in the standard model is a fixed, nondynamical background structure, which does not interact with the quantum fields propagating on it. This Minkowski space nevertheless plays an important role, since the unitary representations of its group of isometries, the Poincar{\'e} group, determine the possible particle content of the quantum field theories defined on it. 

If one does not rely on a perturbative expansion and a distinguished Minkowskian background metric, 
there is no obvious or unique way to construct a theory of quantum gravity that has general relativity as its classical limit. Over time, this has led to a proliferation of approaches, based on ingredients and principles that sometimes deviate significantly from those of classical gravity. All of them have to deal with major technical and conceptual difficulties related to the field content and symmetry structure of gravity. These include an appropriate implementation of diffeomorphism invariance in the quantum theory and an adaptation of standard quantum field-theoretic methodology to take into account the dynamical nature of spacetime geometry. In particular, also ``time'' is a dynamical concept and -- like in general relativity -- there is no notion of time that is distinguished a priori. 

Popular accounts of quantum gravity often characterize this state of affairs as due to the incompatibility between the fundamental principles of quantum theory and general relativity. 
This is a potentially misleading statement, inasmuch as it refers to the absolute, Newtonian time of quantum mechanics (a theory of limited applicability, which is not even compatible with special relativity) and seems to suggest that there is something wrong with the fundamental tenets of either quantum field theory (Hilbert space, superposition principle, unitarity, locality, etc.) or those of general relativity. However, nothing of this kind has ever been shown. On the contrary, toy models of two-dimensional quantum gravity, over which one has excellent analytical control \cite{David1993,Ambjorn1998}, demonstrate that dynamical geometry and quantum theory do not contradict each other as a matter of principle. This does not mean that the principles of quantum (field) theory are sacrosanct, but abandoning them at the Planck scale will usually pose big problems when trying to recover them on macroscopic scales, where they are known to hold. A similar kind of difficulty appears when a chosen Planckian set-up is very far removed from the geometric ingredients of the classical theory (cf.\ Q13 below). 
For example, in formulations based on fundamentally discrete and structureless quantum ``bits'', which lack basic notions like dimensionality, causality or distance, it tends to be very difficult to show that gravity as we know it emerges on sufficiently large scales.

\section*{Q4: Are there specific properties of classical gravity that hamper its quantization?}
\label{sq4}
\addcontentsline{toc}{section}{Q4: Are there specific properties of classical gravity that hamper its quantization?}

As already indicated in Q3, the answer is ``yes''; gravity differs in several ways from the other fundamental interactions,  and requires new ideas for constructing its quantum counterpart, or at least a generalization of the standard methodology of relativistic quantum field theory. Apart from being a weak force, classical gravity is distinguished by its universality: all forms of matter and energy interact gravitationally and are themselves sources of gravity. This leads to the picture of gravity as a local metric structure \emph{of} spacetime, rather than \emph{on} spacetime. There is no a priori fixed, flat background metric and no associated global Lorentz and Poincar{\'e} symmetry, unless one restricts consideration to weak-field perturbation theory. Note also that the invariance group of general relativity, the group of four-dimensional diffeomorphisms, is not a group of local gauge transformations. Unlike the other fundamental interactions contained in the standard model, gravity is therefore not a theory of Yang-Mills type.     

To what extent the differences between gravity and the other forces are significant, and what this implies for its quantum theory are age-old questions, whose answers depend on time and which part of the quantum gravity community is consulted. The traditional particle physics point of view is that (perturbative) quantum gravity is ``just another relativistic quantum field theory'', 
with advocates of grand unification in particular stressing the commonalities of all the forces.
By contrast, general relativists tend to argue that the split 
\begin{equation}
g_{\mu\nu}(x)=\eta_{\mu\nu}+h_{\mu\nu}(x)
\label{split}
\end{equation}
of the spacetime metric $g$ into a Minkowski background $\eta$ and a small perturbation $h$ 
is incompatible with the geometric and nonlinear character of Einstein's theory, and that a proper formulation of quantum gravity should
be background-independent.   
Note that any other, non-flat solution to the Einstein equations can in principle serve as a background, for example, the metric of a spatially homogeneous and isotropic FLRW universe. The propagation of linearized gravitational fields on such a background can be described in the framework of {\it quantum field theory in curved spacetime} \cite{Fewster2012,Brunetti2013}. 

Lastly, the classical theory of general relativity is not exactly simple in terms of its field content, invariances and dynamical equations, which one may expect to see reflected in the quantum theory as well. The complete curvature content is captured by the 256 components of a rank-4 tensor, the Riemann curvature tensor $R^\mu{}_{\nu\rho\sigma}(x)$, which depends on the Lorentzian four-metric $g_{\mu\nu}(x)$, a symmetric, nondegenerate rank-2 tensor of indefinite signature $(-++\ +)$, and its first and second derivatives. The Einstein equations 
\begin{equation}
R_{\mu\nu}[g,\partial g,\partial^2 g]-\frac{1}{2} g_{\mu\nu} R[g,\partial g,\partial^2 g]= 8 \pi G_{\rm N} T_{\mu\nu}[\Phi]
\label{einstein}
\end{equation}
are a set of 10 coupled nonlinear partial differential equations for the metric, which must be solved for given boundary conditions and a given matter content $\Phi$, specified by the energy-momentum tensor $T_{\mu\nu}[\Phi ]$ \cite{Choquet2008}. In the presence of a
cosmological constant $\Lambda$, one adds the term $g_{\mu\nu} \Lambda$ to the left-hand side or, equivalently, includes minus this
term in $T_{\mu\nu}$ on the right-hand side.

Gravity is self-interacting, i.e.\ nonlinear even when $T_{\mu\nu}$ vanishes, and the general solution to eq.\ (\ref{einstein}) is not known. The exact solutions that have been found have special geometric or algebraic properties \cite{Stephani2009}. To solve the Einstein equations in a strong-gravity regime, say, in the vicinity of a pair of inspiralling black holes, one must rely on the methods of numerical relativity \cite{Baumgarte2021}.

\section*{Q5: Are the problems posed by these specific properties merely \textit{\textbf{technical}}?}
\label{sq5}
\addcontentsline{toc}{section}{Q5: Are the problems posed by these specific properties merely \textit{\textbf{technical}}?}

Some of them certainly are technical, and are being addressed. There is good progress in adapting quantum field-theoretic methods to gravity in a way that is background-independent, manifestly or otherwise. However, taking the dynamical nature of spacetime seriously beyond weak-field perturbation theory raises issues that go beyond mere technicalities. For example, 
part of the standard tool box for dealing with the infinities of a relativistic quantum field theory are renormalization techniques, which are used to determine the behaviour of the theory's parameters as a function of scale. 
However, in a theory where geometry is dynamical and no background is distinguished a priori, there also is no pre-existing scale \cite{Ambjorn2020}. If all goes well, such a scale or yardstick may become available as part of a \emph{quantum spacetime}, i.e.\ a solution to the quantum-gravitational dynamics. In any given candidate theory of quantum gravity it is rather nontrivial to show that a notion of scale emerges dynamically, and that renormalization group methods can be meaningfully applied. It is encouraging that blueprints for either mechanism exist in a nonperturbative context \cite{Ambjorn2014}. However, subtle issues on the interpretation of ``scale'' (cf.\ Q23) and on which quantities should be kept fixed during renormalization remain to be understood in greater generality. 

There is another feature of gravity where technical and conceptual difficulties are intertwined at the classical level and even more so in the quantum theory. They have to do with the invariance of gravity under diffeomorphisms, 
and are related to how curved spacetimes are described in general relativity. Mathematically speaking, classical spacetimes are modelled by differentiable manifolds $M$ endowed with a Lorentzian metric $g_{\mu\nu}(x)$ \cite{Choquet1982}. This is both convenient and powerful, since it allows us -- through the use of local coordinate charts $\{ x^\mu \}$ -- to do computations just like on $\mathbb{R}^4$. The choice of coordinates is arbitrary; the laws of tensor calculus tell us how the metric, the Riemann tensor and other relevant quantities transform under a change of coordinates.
For example, under a coordinate transformation $x\mapsto y(x)$, the metric $g_{\mu\nu}(x)$ will assume the form $\tilde{g}_{\rho\sigma}(y)$, where
\begin{equation}
\tilde{g}_{\rho\sigma}(y)=\frac{\partial x^\mu(y)}{\partial y^\rho} \frac{\partial x^\nu (y)}{\partial y^\sigma}\ g_{\mu\nu}(x(y)).
\label{gtrans}
\end{equation}
Despite the fact that the tensor components of the metric $\tilde{g}$ have in general a different functional form from those of $g$, both are physically equivalent; they describe the same curved spacetime geometry in terms of different coordinates. The situation is analogous to that in gauge field theory: the classical description is redundant, since there are nontrivial transformations of the fields that leave the physics unchanged. 

We do not usually worry about this redundancy in general relativity. On the contrary, we may use the flexibility in choosing coordinates to bring a {\it given} metric into a particularly simple form, e.g.\ one that is adapted to its isometries. Nevertheless, in interpreting any results it is important to distinguish between ``physics'' and ``gauge'', i.e.\ between what is (potentially) observable and what is a mere coordinate effect. This is not entirely straightforward, as illustrated by some infamous examples from the history of general relativity, 
including the long-standing confusion about the physical status of the event horizon at $r\! =\! 2 G_{\rm N} M$ of the Schwarzschild metric, or the dispute about whether gravitational waves are a true physical phenomenon \cite{Kennefick1997}. 

When it comes to the quantum theory, the classical ``freedom to choose coordinates'' becomes a major source of problems. The diffeomorphism group must be represented on quantum states without picking up anomalies and in a way that is compatible with renormalization. The associated gauge degrees of freedom lead to a potential overcounting and therefore infinities in the path integral. To deal with them, gauge-theoretic methodology involving gauge-fixing and Faddeev-Popov determinants must be adapted to gravity  \cite{Mottola1995} and to a nonperturbative context, also taking into account the nonlinearities of the physical configuration space. Finally, invariant quantum observables must be identified to describe the true physics of quantum gravity.

\section*{Q6: What are observables in gravity?}
\label{sq6}
\addcontentsline{toc}{section}{Q6: What are observables in gravity?}

In close analogy to gauge field theory, observables in classical gravity are those functions on its configuration space $\text{Lor}(M)$, 
the space of all Lorentzian metrics on a manifold $M$, which are invariant under the action of the diffeomorphism group
$\text{Diff}(M)$. The corresponding physical configuration space is sometimes called the {\it space of geometries}, 
and can be represented by the quotient
\begin{equation}
\mathcal{G}(M):= \frac{\text{Lor}(M)}{\text{Diff}(M)},
\label{quotient}
\end{equation}
whose elements are equivalence classes $[g_{\mu\nu}(x)]$ of Lorentzian metrics under diffeomorphisms. In other words,
observables are functions on the space $\mathcal{G}(M)$.

Let us be a bit more specific about the mathematical nature of these structures.
A (global) diffeomorphism is a smooth ($C^\infty$) map $\varphi: M\! \rightarrow\! M$ with smooth inverse map $\varphi^{-1}$. 
Diffeomorphisms play the role of structure-preserving maps of differentiable manifolds, where two manifolds $M$, $M'$ that are related by a 
diffeomorphism are regarded as equivalent.
The diffeomorphisms of $M$ form an infinite-dimensional group $\text{Diff}(M)$ under composition. 
In local coordinate charts, the components of tensors transform under diffeomorphisms in the usual way; 
the transformation behaviour of the metric two-tensor in eq.\ (\ref{gtrans}) above is a special case.
\begin{figure}[t]
\centerline{\scalebox{0.65}{\rotatebox{0}{\includegraphics{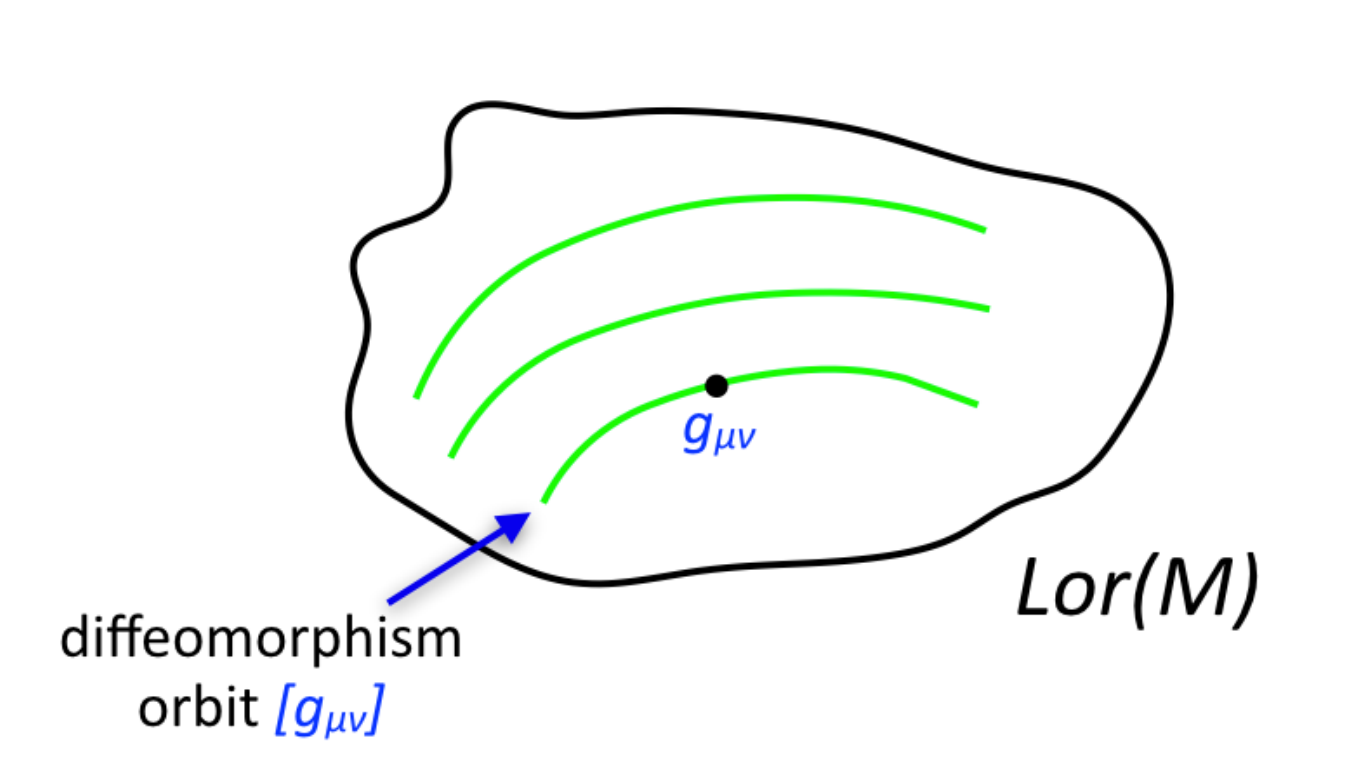}}}}
\caption{Gauge orbit through a Lorentzian metric $g_{\mu\nu}\in \text{Lor}(M)$ under the action of $\text{Diff}(M)$.}
\label{orbit}
\end{figure}

Two Lorentzian manifolds that are related by a diffeomorphism represent the same geometry and therefore the same physics,
which means that they are physically indistinguishable. Invoking a concept familiar from gauge theory, for a given Lorentzian metric $g$, corresponding
to a point in $\text{Lor}(M)$, we can consider the {\it gauge orbit} $[g]$ through $g$, consisting of all metrics $g'=\varphi\cdot g$ 
that are related to $g$ by a diffeomorphism $\varphi$ (Fig.\ \ref{orbit}). For generic metrics $g$, which do not possess any isometries, the gauge
orbits are all of the same size, and correspond to points in the (infinite-dimensional) physical configuration space $\mathcal{G}(M)$.
As mentioned before, observables can simply be characterized as functions on $\mathcal{G}(M)$.  

However, the concrete identification of such observables in pure gravity, in the 
absence of matter and other background or reference structures, is not completely straightforward. Firstly, the action of $\text{Diff}(M)$ on $\text{Lor}(M)$ (beyond the linearized case)
is complicated and the quotient space is a nonlinear and not particularly nice space\footnote{It should be noted that also $\text{Lor}(M)$ is 
not a linear space, because of the inequality $\det g <0$.}. This makes it challenging to parametrize the space $\mathcal{G}(M)$ 
of geometries in terms of a set of
nonredundant field variables, which is particularly desirable from the point of view of the quantum theory. Fortunately, there is a powerful and
gravity-specific way to address this issue in quantum gravity, based on {\it Regge geometries}, which will be described in Q26 below.

Secondly, the fact that pure gravity is diffeomorphism-invariant is basically tantamount to saying that the theory is {\it background-independent}\footnote{see \cite{Giulini2006} for a detailed discussion of the relation between diffeomorphism invariance and background independence and the
subtleties involved}, which in turn implies that
its observables are nonlocal. This can be understood in various ways, for example, by noting that a diffeomorphism can
be regarded as actively moving points on the manifold. This is not the case in gauge field theory, where gauge 
transformations act locally, typically amounting to a rotation in some internal linear space at each point $x\in M$, without affecting the underlying
manifold.
The deeper reason for the nonlocality of gravitational observables is the dynamical nature of the geometric structure (which at the same time is the structure we use to {\it locate} whatever happens in spacetime): since the metric always gets ``moved along''
when we apply a diffeomorphism, ``the point $x$'' is not an invariant concept. As a consequence, the field $\phi (x)$ at the point $x$ is not
a diffeomorphism-invariant quantity and therefore not a physical observable.   
By contrast, simple examples of (nonlocal) observables are given by spacetime integrals of scalar quantities, like the total volume of $M$ or the 
integral of the Ricci scalar $R(x)$,
\begin{equation}
\int_M d^4 x\, \sqrt{|\det(g)|}\, R(x),
\label{ricciscalar}
\end{equation}
both of which are part of the Einstein-Hilbert action $S^{\rm EH}$ in the presence of a cosmological constant $\Lambda$,
\begin{equation}
S^{\rm EH}[g]=\frac{1}{16\pi G_{\rm N}}\, \int_M d^4 x\, \sqrt{|\det(g)|}\, (R -2 \Lambda ),
\label{SEH}
\end{equation}
which governs the dynamics of general relativity in vacuum. 
Other examples are diffeomorphism-invariant $n$-point functions (see \cite{Ambjorn1996} for an illustration in two dimensions). 
Anticipating developments in the quantum theory that will be
discussed below, the technical complications arising from the presence of unphysical gauge degrees of freedom in gravity are in
principle familiar from (nonabelian) gauge field theories. One can try to address them by conventional gauge-fixing methods in the continuum 
or by alternative parametrizations of the space $\mathcal{G}(M)$ (or regularized versions thereof). 
However, the strongly nonlocal character of observables is unusual
and sets gravity apart from the other interactions. This should be viewed as a fundamental and inherent feature of the quantum
theory, which requires adapted tools and treatment, and at the same time determines what nonperturbative quantum gravity and quantum spacetime are all about. 
As we will see in Q28 and Q29 below, nonlocal quantum observables can be constructed in nonperturbative quantum gravity and are already providing valuable information about the quantum theory.

\section*{Q7: Why should I learn about perturbative quantum gravity?}
\label{sq7}
\addcontentsline{toc}{section}{Q7: Why should I learn about perturbative quantum gravity?}

While the focus of these lecture notes is on {\it non}perturbative quantum gravity, we will in Q7--Q10 also comment 
on the perturbative approach,
by which the subject was addressed traditionally. The first attempt to quantize the linearized gravitational field is due to Bronstein
\cite{Bronstein1936}. In a more modern language, the key quantity of interest is the gravitational path integral, formally defined as
\begin{equation}
Z=\int\limits_{{\cal G}(M)}\!\! {\cal D} [g]\, {\rm e}^{i S^{\rm EH}[g]},
\label{pathint}
\end{equation}
a quantum superposition of equivalence classes $[g]$ of metrics, each contributing with a complex amplitude depending on
the gravitational action (\ref{SEH}). In perturbation theory one decomposes the metric according to eq.\ (\ref{split}), expands
the action into a ``free'' part quadratic in the linearized fields $h_{\mu\nu}$ plus terms of higher order, and performs the
functional integral (\ref{pathint}) over the equivalence classes $[h]$ (under the action of the linearized diffeomorphism group)
instead of $[g]$. The perturbative expansion in terms of Feynman diagrams 
is described for instance in \cite{DeWitt1967,Veltman1975}. 

The subtext of the question above is the fact that gravity is not perturbatively renormalizable and the standard methods of perturbative
quantum field theory on a Minkowskian background are therefore not sufficient to construct a fundamental theory of quantum gravity that is 
valid on all scales. Nevertheless, the claim is that there are quantum effects of gravity which can be computed unambiguously
when using the perturbative framework in the sense of an {\it effective field theory}, valid on energy scales that are small compared to the Planck mass \cite{Burgess2003,Donoghue2012}. A well-known example are quantum corrections to the Newtonian potential \cite{Khriplovich2002,BjerrumBohr2002}, which share the characteristic feature of such corrections in being extremely small and 
far out of any experimental reach, certainly within a solar system setting. 
This raises questions about the status and physical relevance of the perturbative sector of quantum gravity,
even though it circumnavigates the issue of nonrenormalizability. 

Nevertheless, it is worthwhile to get acquainted with the toolbox and results of perturbative quantum gravity.
At the very least, one can
appreciate their sheer computational complexity
and why it took so long to identify all contributions to the quantum corrections to Newton's potential, or to
establish the nonrenormalizability of pure gravity at two-loop order \cite{Goroff1985,vandeVen1991}, say. 
Interestingly, it has been shown in more recent times that such computations can be simplified to some extent by using
relations between the scattering amplitudes of gauge theory and gravity in the weak-coupling limit 
\cite{Bern2002}, see \cite{BjerrumBohr2013,Bern2015} for some explicit applications.
The difficulty of separating true observables from coordinate effects persists in perturbative gravity. Addressing the
associated issues of gauge dependence and renormalization is in general subtle, as is illustrated by cosmological applications,
see \cite{Frob2017} for a recent example. Identifying suitable cosmological (quantum) observables could serve as a stepping
stone towards nonperturbative quantum gravity and provide inspiration and possible points of comparison. However, apart from
dealing with the intrinsic difficulties of both the perturbative and the nonperturbative approaches, this will require 
establishing common ground (in particular, common observables) between formulations with and without a fixed background.

\section*{Q8: What are gravitons?}
\label{sq8}
\addcontentsline{toc}{section}{Q8: What are gravitons?}

Gravitons are hypothetical massless spin-2 particles, which are the force carriers of the linearized weak gravitational 
field $h_{\mu\nu}(x)$ on a Minkowskian background, in accordance with the perturbative split \eqref{split}. They are the gravitational
analogues of photons, the massless spin-1 particles that are the force carriers of the electromagnetic field 
and are described by the gauge potential
$A_\mu (x)$. Gravitons are hypothetical in the sense that individual gravitons have not been observed and heuristic arguments have been 
put forward that this may never happen, as a matter of practice or even principle \cite{Rothman2006,Dyson2013}. 
On the one hand, the culprit is again the weakness of gravity, requiring graviton detectors so massive that they would collapse 
into black holes or waiting
times of the order of the lifetime of the universe. On the other hand, it is conceivable that gravitons -- unlike photons -- do not exist as individual, detectable particles, despite having a similar theoretical status as ingredients of an (effective) perturbative relativistic quantum field theory. Even if this was the case, it does not imply that a quantum theory of gravity does not exist. 
For example, staying in a weak-field setting, it has been
suggested that the quantum nature of gravity may be demonstrated by looking for gravitationally induced
entanglement, for example, in quantum superpositions of massive particles \cite{Bose2017,Marletto2017}. 
There is currently a flurry of activity in this area and an ongoing debate about what, if anything, can be learned from these suggested 
tabletop experiments about gravitons and, more generally, about the true degrees of freedom of quantum gravity \cite{Anastopoulos2018,Carney2021,Danielson2021}. This is a fascinating discussion, which may take some time to settle.  

Meanwhile, let us sketch the perturbative Fock space construction where gravitons make an appearance. Recall that
the presence of a unique vacuum and the associated particle picture are intimately tied to the presence of a 
Minkowskian background metric $\eta_{\mu\nu}$, on which the 10-parameter Poincar{\'e} group of spacetime 
translations, rotations and boosts acts as a global symmetry group. Quantum particles and fields on Minkowski space
are classified according to their mass and spin, which in turn characterize unitary irreducible representations of the Poincar{\'e} group. 

The close analogy with electromagnetism can be exhibited by working with
the linearized weak fields $h_{\mu\nu}(x)$ in harmonic gauge, imposing the gauge condition 
$\partial_\rho h^\rho{}_\mu-\frac{1}{2} \partial_\mu h^\rho{}_\rho =0$. 
The linearized vacuum Einstein equations then take the form
\begin{equation}
\Box\, h_{\mu\nu}(x)=0,
\label{weakeom}
\end{equation}
whose solutions are plane waves. This is the counterpart of the wave equation
$\Box\, A_\mu =0$ for the electromagnetic potential in Lorenz gauge ($\partial_\rho A^\rho=0$)
in the absence of sources.
In transverse-traceless gauge \cite{Carroll2004}, the wave solutions to eq.\ (\ref{weakeom}) can be written in matrix form as
\begin{equation}\label{gravwave}
    h_{\mu\nu}=
  \,  \begin{pmatrix}
        0 & 0 & 0 & 0  \\
        0 & a & b & 0 \\
        0 & b & -a & 0 \\
        0 & 0 & 0 & 0
    \end{pmatrix}\,
    e^{i\omega_{\vec k} (z-t)}, 
\end{equation}
where $\omega_{\vec k}\! =\! |{\vec k}|$ is the wave's frequency, $a,b \!\in\! \mathbb{R}$ are the amplitudes of the wave's so-called 
plus- and cross-polarizations
and for definiteness we have chosen a wave vector $\vec k$ along the positive $z$-direction. A gravitational
wave therefore has two linearly independent polarizations, like an electromagnetic wave, but they behave differently 
under spatial rotations (in the plane perpendicular to the wave
vector): a rotation by $180^{\degree}$ acts on the polarizations as the identity transformation, whereas for electromagnetic waves this is only the case for a $360^{\degree}$-rotation.

The properties of the classical field $h_{\mu\nu}$ translate in a straightforward way to properties of the field operators $\hat{h}_{\mu\nu}(x)$ 
in the quantization of linearized gravity, whose bosonic Fock space is built from $n$-graviton states. The one-particle states 
$|\vec{k},\sigma\rangle$ transform in the unitary irreducible representation of mass 0 and spin 2 of the Poincar{\'e} group, and are
labelled by the three-momentum $\vec k$ and the helicity $\sigma =\pm 2$, counterpart of the classical polarization. The corresponding
creation and annihilation operators $\hat{a}^\dagger (\vec{k},\sigma)$ and $\hat{a} (\vec{k},\sigma)$ appear in the expansion of
the field operator,
\begin{equation}
\hat{h}_{\mu\nu}(x)=\int\frac{d^3k}{(2\pi)^{3}} \frac{1}{\sqrt{2 \omega_{\vec k}}}\sum_{\sigma=\pm 2} \big[ \hat{a}(\vec{k},\sigma)
\epsilon_{\mu\nu}(\vec{k},\sigma){\rm e}^{ik\cdot x}  +  \hat{a}^\dagger(\vec{k},\sigma)
\epsilon^\ast_{\mu\nu}(\vec{k},\sigma){\rm e}^{-ik\cdot x}    \big],
\label{fieldop}
\end{equation}
where $\epsilon_{\mu\nu}$ denotes the polarization tensor. This again underlines the similarities between linearized gravity and electromagnetism, up to the more complicated tensorial structure of the former (two indices rather than one).\footnote{A detailed
comparison between electromagnetism and general relativity, including their properties under renormalization, can be found in
\cite{Woodard2009}.} Of course, one should keep in
mind that the classical linearized theory -- which is the starting point of perturbative quantum gravity -- approximates Einstein's theory 
only when fields are weak. This excludes many interesting nonlinear strong-gravity phenomena, which is indicative of 
the limited range of applicability of the perturbative theory in the quantum realm.

\section*{Q9: What is the meaning of ``nonrenormalizable''?}
\label{sq9}
\addcontentsline{toc}{section}{Q9: What is the meaning of ``nonrenormalizable''?}

It has been known for a long time, and had been anticipated even longer, that gravity is perturbatively 
nonrenormalizable. This 
means that the divergences which appear when evaluating momentum integrals at higher orders in the perturbation 
expansion \`a la Feynman cannot be absorbed by renormalizing a finite number of coupling constants. 
In the case of gravity, the coupling constants are associated with an infinite tower of
higher-order curvature terms, which for consistency must be added to the Einstein-Hilbert Lagrangian during renormalization. 
Their contribution to the low-energy physics is negligible, but more and more of them become important as the Planck 
scale is approached. Since each renormalized coupling constant needs to be fixed by experiment and since at the Planck scale
we would need to fix infinitely many such parameters, the perturbative theory has no predictive power at this scale. This is what lies behind
the statement that perturbative quantum gravity cannot serve as a fundamental theory of quantum gravity. Of course, given the 
dearth of quantum gravity experiments and observations, in practice such an approach runs into difficulties long 
before the Planck scale is reached.  

As is explained in many textbooks on the subject, the occurrence of divergences is a standard feature of 
relativistic quantum field theories and 
for the case of renormalizable theories can be dealt with by standard methods (see e.g.\ \cite{Zee2010}). 
A simple criterion for whether or not a theory may be renormalizable involves the mass dimension of its coupling constant. When
it is negative, a theory is said to be ``power-counting nonrenormalizable''. This is the case for gravity, where the
mass dimension of Newton's constant $G_{\rm N}$ is $-2$. A heuristic way of seeing why this may cause a problem
is to consider quantum corrections to some gravitational observable which at lowest order is given by $\mathcal O$. 
To regularize any momentum integrals, let us work with a
large momentum cutoff $\Lambda$. The corrections come in the form of a power series in $G_{\rm N}$, which for
dimensional reasons will look like 
\begin{equation}
{\cal O}\,\big( 1+(\dots)G_{\rm N}\Lambda^2+(\dots)(G_{\rm N}\Lambda^2)^2+\dots\big),
\label{obexp}
\end{equation}
where $(\dots)$ indicates numerical coefficients. As the regulator is removed, $\Lambda\rightarrow \infty$, the correction
terms diverge, at a rate that gets worse with increasing order. The resulting ultraviolet infinities affect computations of quantities
like graviton scattering amplitudes, rendering their physical status rather unclear. 

To determine the renormalizability of gravity in perturbation theory, one investigates the (effective) gravitational Lagrangian order by
order and determines the numerical coefficients that multiply the potentially divergent terms (see \cite{Kiefer2012} for
further details and references). It turns out that the
contribution at one-loop order vanishes in pure gravity (although not in the presence of matter) \cite{tHooft1974}. However, 
gravity is not finite at two-loop order, due to a term cubic in the curvature whose coefficient is divergent \cite{Goroff1985,vandeVen1991}.
Since there are potentially divergent terms at all higher loop orders and since
the vanishing of coefficients in the perturbation series is usually related to the presence of some symmetry, there is
no reason to believe that these contributions will be finite. Hence one concludes that gravity is perturbatively nonrenormalizable. 
As pointed out above, the perturbative formulation may still be useful as an effective framework at energies much smaller than 
the Planck scale, but constructing a fundamental theory of quantum gravity clearly needs other tools and ideas.

\section*{Q10: Why does perturbative quantum gravity fail?}
\label{sq10}
\addcontentsline{toc}{section}{Q10: Why does perturbative quantum gravity fail?}

There are about as many answers to this question as there are ideas of how to address the failure of perturbative 
quantum gravity. To understand this landscape of ideas, it is useful to know that there is a traditional split between 
the viewpoints of particle physicists and general relativists, 
which has had ramifications for the different types of quantum theory that have been considered in the wake of this negative result. 

Loosely speaking,
the particle physics perspective emphasizes the structural similarities between gravity and other (gauge) field theories
and at the same time relies as much as possible on the power of perturbation theory. Trying to evade the perturbative
nonrenormalizability naturally leads to modifications
of general relativity, e.g.\ by adding higher-derivative terms to the action \cite{Stelle1976} or by 
modifying and enlarging its field content and symmetries, 
culminating in grand unified theories, including superstring theories.  
By contrast, from a general relativistic perspective the dynamical nature of spacetime sets gravity apart from the other interactions, 
and the perturbative decomposition (\ref{split}) of the metric is incompatible with the nonlinear, geometric structure of 
the physical configuration space ${\mathcal G}(M)$ and the Einstein equations \cite{Isham1993}. The emphasis is therefore
on developing 
nonperturbative methods that can accommodate these aspects, either in a canonical operator quantization -- whose most recent
incarnation is loop quantum gravity (see Q21) --
or by working with alternative parametrizations of the space ${\mathcal G}(M)$, like those based on Regge geometries \cite{Regge1961}.  
Also here one hopes to find a natural resolution of the perturbative infinities. -- 
Although many of the ideas and approaches of either ``flavour'' are not necessarily mutually exclusive, they often look sufficiently 
different in practice to obscure possible communalities and complementarities, in addition to being incompletely developed.  

Other points of criticism of perturbative quantum gravity are its lack of background independence, given its reliance on the underlying
Minkowski space for much of its construction.\footnote{One may of course consider background metrics other than the Minkowski metric, 
but these are associated with additional difficulties including the generic nonuniqueness or complete absence of a vacuum state, 
and an associated S-matrix formulation \cite{Hollands2014}.} Furthermore, many practitioners have doubts 
about the suitability of metric fields to describe 
a strongly quantum-fluctuating Planckian spacetime regime, leading to more or less radical suggestions for alternatives, 
from fundamentally discrete (sub-)Planckian ``building blocks'' to other ``ultraviolet completions'', 
where completely different, non-metric continuum degrees of freedom take over at the Planck scale. 
Lastly, there may be nothing wrong with perturbation theory as
such, but we could be perturbing around the ``wrong'' vacuum. Perhaps it can be shown that gravity is {\it nonperturbatively
renormalizable} \cite{Weinberg1979} (see also Q23), which raises the question of how such an alternative vacuum state can be 
found nonperturbatively.

\section*{Q11: Can we leave gravity unquantized?}
\label{sq11}
\addcontentsline{toc}{section}{Q11: Can we leave gravity unquantized?}

The claim that gravity is the only fundamental interaction that is essentially classical, without an underlying quantum theory, 
resurfaces every so often. The attention this idea seems to generate,
mainly outside of theoretical high-energy physics, is somewhat out of proportion with its scientific merit. 
It lacks basic plausibility because it runs counter to several fundamental physical principles
that have been discovered during the 20th century. First and foremost, we have learned that the physical world at a fundamental,
microscopic level is governed by {\it quantum} laws. This is true for all forms of matter and their interactions which have been studied
experimentally. In other words, the quantum nature of the theories describing them reflects a physical reality.
What we think of as ``classical physics'' 
is only an approximation, valid at sufficiently low energies and on sufficiently large scales, to a more accurate description that is
inherently quantum. The other key lesson, coming from general relativity, is that gravity is {\it universal}, in the sense that all forms of matter and
energy interact gravitationally. Literally everything serves as a source of gravitational forces and 
in turn is subject to those forces, all of which is encoded into the local curvature structure of spacetime. 
Given the universality of both quantum theory and gravitation, there is 
little basis for conjecturing that the interaction of quantum matter and gravity is not governed by quantum laws or that
a quantum theory of gravity does not exist.

The quantum nature of gravity needs to be confirmed by experiment or observation; it cannot be proven by logical or
mathematical deduction. The fact that we have not yet seen quantum-gravitational effects does not provide strong evidence 
in favour of fundamentally classical gravity, but is explained perfectly well by the weakness of gravity, as outlined in Q3 and Q7.
Supporters of the idea that gravity should not be quantized should not only provide a convincing motivation, but also 
a theoretical framework in which classical gravity -- viewed as a fundamental theory -- is coupled 
consistently to quantum matter. One type of approach is based on the so-called semiclassical Einstein equations,  
\begin{equation}
R_{\mu\nu}-\frac{1}{2}\, g_{\mu\nu} R=8\pi G_{\rm N}\, \langle\Psi |\, \hat{T}_{\mu\nu}\, |\Psi\rangle,
\label{semicl}
\end{equation}
where geometry (in the form of the left-hand side of the Einstein equations (\ref{einstein})) is coupled to the expectation value of the 
quantized energy-momentum tensor ${\hat T}_{\mu\nu}$ in some quantum state 
$|\Psi\rangle$.\footnote{The idea of treating semiclassical gravity as fundamental should be distinguished from its use as 
an approximation and intermediate step towards 
quantum gravity, where -- unlike in quantum field theory on a fixed curved manifold -- the quantum fields
can back-react on the spacetime geometry via (\ref{semicl}), see e.g.\ \cite{Wald1994}.}
Apart from technical difficulties in solving (\ref{semicl}) alongside dynamical equations for the quantum matter fields, which also
depend on the metric $g$, it is currently unclear whether this or other approaches to semiclassical gravity can be fleshed out
and turned into self-consistent fundamental theories, see \cite{Kiefer2012,Grossardt2022} for further discussions and references.
Equally unclear is whether and how the singularities of general relativity would be resolved in such a framework, something
that is usually assumed to be taken care of by a quantum theory of gravity.

\section*{Q12: Must quantum gravity be part of a unified theory of all interactions?}
\label{sq12}
\addcontentsline{toc}{section}{Q12: Must quantum gravity be part of a unified theory of all interactions?}

A complete, theoretical description of all fundamental forces will also contain quantum laws for how gravity and matter
(or whatever becomes of these notions at the Planck scale) interact with each other. It is not known whether there is a 
single dynamical or symmetry principle which governs fundamental physics at very high energies and unifies all of the known forces;
there is currently no compelling evidence for or against such a scenario.    
Following the successes of quantum gauge field theories and the standard model, the search for a unified theory became a 
prominent theme in particle physics research, whose ultimate goal -- beyond a grand unification of the forces of the standard
model -- is also the incorporation of gravity. The fact that this aim is proving difficult to achieve could be an indication that the
concept of unification has reached the limits of its usefulness or that gravity is sufficiently different from the other interactions 
to defy unification, or both.

A prominent example of a unified theory is superstring theory \cite{Polchinski2005,Mohaupt2022}, 
which is based on a unified dynamical principle of tiny vibrating strings in a ten-dimensional spacetime. 
However, the scope and perspective of string theory have changed significantly over the last decades, 
from that of a candidate for an essentially unique ``theory of everything'', 
including quantum gravity, to a theoretical framework for addressing a variety of physics problems in novel ways,
from cosmology to condensed matter systems. 
The richness of string theory (in terms of degrees of freedom, symmetries and extra dimensions) 
and the ensuing high degree of nonuniqueness stand in the way of a fundamental interpretation, as does the absence of 
a nonperturbative definition of the theory (often captured by the notion of M-theory, see \cite{ncatlab} for an annotated literature guide). 
Although string theory has gravitonic excitations in its spectrum of vibrational states, it 
has arguably not added much to our understanding of nonperturbative quantum gravity in four dimensions, for example, 
by illuminating the nature of the theory and its observables in a Planckian regime. This is also true for more recent
developments in the context of the AdS/CFT correspondence \cite{Ammon2015}. One should also recall that it is not
straightforward to incorporate a positive cosmological constant into string theory, a property that seems to be required by
observation.   
 
Lastly, let us comment on the idea that considering quantum gravity without simultaneously including some form of
quantum matter may be meaningless or inconsistent, even in the absence of a grand unifying principle including gravity.\footnote{We are
leaving aside questions about the nature of the observer, which are more pertinent in the context of cosmology.}
There is currently little evidence to support this view. 
Studies of ``pure'' quantum gravity, without adding any matter by hand,
are the quantum analogue of solving the vacuum Einstein equations (where the energy-momentum tensor
$T_{\mu\nu}$ vanishes, possibly up to a cosmological-constant term, which can be included on the left-hand side of the equations). 
These classical equations possess many nontrivial solutions, including the generation of black
holes by the collision of gravitational waves \cite{Christodoulou2008}. 
There is no obvious reason why such solutions should not have well-defined quantum counterparts.
Investigating the nonperturbative dynamics of pure quantum gravity is well motivated and interesting in its own right, and
in addition serves as a starting point
to systematically study the influence and consistency of specific quantum matter inhabiting quantum spacetime, 
see e.g.\ \cite{Dona2013}. 
This strategy has been very fruitful in the analysis of nontrivial toy models of nonperturbative quantum gravity in 
two dimensions, both without and with matter coupling, cf.\ Q17. 

It is a dynamical question whether and on what length scale gravity and matter are strongly coupled, and  
which degrees of freedom are dominant at the Planck scale. 
Among the conceivable scenarios are a complete dominance of the Planckian dynamics by some kind of
quantum-geometric excitations, perhaps even including a quantum-gravitational origin of matter.
Alternatively, matter may interact strongly with spacetime at the Planck scale, giving rise to a combined
gravity-matter vacuum state and dynamics.

\section*{Q13: Which a priori choices are required to construct a theory of quantum gravity?}
\label{sq13}
\addcontentsline{toc}{section}{Q13: Which a priori choices are required to construct a theory of quantum gravity?}

This may look like a bookkeeper's rather than a physicist's question, but it is meant to highlight the fact that certain choices {\it have} to be made
a priori when trying to construct a theory of quantum gravity beyond perturbation theory. Making an inventory of the ingredients needed is also helpful in understanding the differences between various approaches.\footnote{see also \cite{Isham1993} for a classification, covering developments until 1993} The choices can be divided broadly into four categories:
\vspace{-0.2cm}

\begin{description}[itemsep=2pt,leftmargin=0pt]
\item[$\bullet$] {\it microscopic, elementary degrees of freedom:} A configuration space must be picked on which the dynamics is defined,
with the standard choice given by some version of the metric or vierbein fields of the classical continuum. 
A popular alternative are discrete building blocks at the Planck scale, serving either as a regulator or representing a form of fundamental discreteness \cite{Surya2019}. A further question is whether these building blocks carry metric information or are non-geometric or
structureless. Does the set-up go beyond standard quantum field theory, by considering fundamental extended objects like strings, loops or higher-dimensional branes? Other degrees of freedom could be dynamical besides the geometric ones, including torsion, topology, 
additional (matter) fields and even dimensionality.
\item[$\bullet$] {\it fundamental principles:} There are a number of fundamental principles, which are corner stones of physics and relativistic
quantum field theory in particular, whose validity at the Planck scale has occasionally been questioned, usually looking for loopholes for why
a particular quantum gravity approach or argument along more conventional lines fails. They include locality, causality and unitarity. Since gravity is a theory of dynamical spacetime, even more fundamental concepts may be called into question microscopically, like the presence of any notion of time or space. One may also hope to discover entirely new general principles that help us to understand quantum gravity. Examples are the existence of a minimal length scale \cite{Garay1994} and the holographic principle \cite{tHooft1993,Bousso2002}.
\item[$\bullet$] {\it symmetries}: The question of which symmetries are present in a given formulation is closely related 
to the choice of the elementary degrees of freedom. One also needs to take into account that it may not be straightforward to
represent a particular symmetry in the quantum theory, for example, because of anomalies. Formulations based on metric or other continuum fields inevitably have to implement some quantum version of diffeomorphism invariance. This means gauge-fixing in covariant 
approaches and solving quantum constraint equations \`a la Dirac in canonical approaches. Other symmetries that may be present
include local Lorentz transformations or other local gauge symmetries (in first-order and other gauge-theoretic formulations of gravity),
supersymmetry (in supergravity and superstring theory) and conformal symmetry (in particular limits).
\item[$\bullet$] {\it dynamical principle}: Lastly, a dynamical principle is required. In other words, what plays the role of the ``quantum Einstein equations''? There are essentially two choices, namely, a nonperturbative version of the gravitational path integral (\ref{pathint}), from which one can compute correlators and other observables, and a canonical quantization, whose key aim is to find physical quantum states $\psi [g]$ which solve the so-called Wheeler-DeWitt equation \cite{DeWitt1967a,DeWitt1997},
\begin{equation}
\hat{\cal H}_\perp \psi[g]:= 
\bigg(-\!\frac{\kappa^2}{2 \sqrt{\det g}} \,\big( g_{ik}g_{jl}+g_{il}g_{jk}-g_{ij}g_{kl} \big)\, \frac{\delta}{\delta g_{ij}}\frac{\delta}{\delta g_{kl}}
-\frac{\sqrt{\det g}}{\kappa^2}\, {}^{(3)}\! R\,\bigg)\,\psi[g]=0,\;\;\;
\label{wdw}
\end{equation}
where $\kappa^2=16 \pi G_{\rm N}$, $g_{ij}$ refers to the Riemannian metric on three-dimensional spatial 
hypersurfaces $\Sigma_t$ of constant time $t$ in a
($3\! +\! 1$)-decomposition of the spacetime metric $g_{\mu\nu}$, and ${}^{(3)}\! R$ is the three-dimensional Ricci curvature scalar, 
see also Q19 and Q20.
\end{description}

\section*{Q14: Is quantum gravity just one big free lunch at the Planck scale?}
\label{sq14}
\addcontentsline{toc}{section}{Q14: Is quantum gravity just one big free lunch at the Planck scale?}

The large range of apparent choices outlined in Q13 reflects the fact that there is no standard methodology
to construct a quantum theory once we leave the realm of perturbation theory. Since the Planck scale is far beyond any
experimental reach, can we let our imagination roam freely when making these choices? For instance,
can we pick our favourite (sub-)Planckian building blocks, assemble them into some kind of quantum spacetimes, put 
those into a path integral, and {\it declare} this to be a theory of quantum gravity? --
Considering the lessons learned from several decades of quantum gravity research, such an approach is unlikely to yield a serious candidate theory, even in the absence of direct probes of Planckian physics.

One potential sticking point is {\it uniqueness}. Recall that we dismissed perturbative quantum gravity on account of an
infinite number of free parameters that must be supplied to extract results at the Planck scale. For quantum gravity to 
be predictive on all scales, it should only have finitely many such parameters, and -- given the current lack of phenomenology -- preferably
as few as possible. If a formalism has many extra degrees of freedom and symmetries, beyond those
motivated directly by general relativity, there is a danger that unwanted parameters proliferate, associated with e.g.\ symmetry
breaking, quantization ambiguities, fine-tuning of couplings or enumerating possible vacuum states. 

Quantum configurations with too many degrees of freedom can also be problematic. 
A case in point is the inclusion of a sum over topologies
in the gravitational path integral, which is sometimes advocated.  
If such a sum is included, the number of states associated with a given spacetime volume grows factorially with the volume, 
which makes the path integral highly divergent. In the simplest case of two-dimensional toy models of Euclidean quantum gravity, one 
can still find generating functions (in the form of certain matrix integrals \cite{Ambjorn1997,Saad2019}) which reproduce the  
topological genus expansion, but they are highly nonunique and require new, physically motivated selection principles. 
In dimension four, there is not even a classification of general topologies. 

Another potential sticking point is the {\it classical limit}: any theory of quantum gravity should at sufficiently low energies and large
distances reproduce key aspects of general relativity, including the presence of four-dimensional spacetimes.\footnote{It is unclear whether it is 
interesting or desirable to reproduce perturbative quantum gravity from the 
nonperturbative theory, because of the unresolved status of the former. Our discussion here focuses on recovering properties of the classical theory.} 
In nonperturbative formulations, identifying and getting a computational handle on the classical limit 
is usually very nontrivial. Models are often insufficiently developed to be able to address the existence or nature of the classical limit.
However, there is a large body of work on statistical models of quantum gravity where classical and continuum limits have been 
studied systematically. Amongst other things, they have uncovered nonperturbative effects that stand in the way of recovering 
extended four-dimensional geometries in a large-scale limit.
These effects are related to particular quantum fluctuations that can entropically dominate path integrals and partition functions at 
short length scales. A notorious pathology of this type are so-called branched polymers, which can be identified
in terms of their critical exponents. Their occurrence appears to be rather generic across dimensions ($\geq\! 3$) and
approaches \cite{Ambjorn1997,Veselov2003,Lionni2017}. A known cure is to require quantum configurations to carry a well-defined 
causal structure, as is done in causal dynamical triangulations \cite{Ambjorn2012} (cf.\ Q27).

The above points illustrate that the construction of a nonperturbative theory of quantum gravity is
significantly constrained even before considering any Planck-scale dynamics or possible phenomenology.  
With regard to giving up principles like causality, locality or unitarity at the Planck scale, the difficulty is then
to show that they are restored dynamically on macroscopic scales. 
Likewise, to demonstrate that spacetime as we know it can be recovered from some 
microscopic quantum substrate -- often of a discrete nature -- is equally challenging 
(see \cite{Major2006,Konopka2006,Kelly2019,Obster2022} for some examples). 

This raises the more general question of which aspects of gravity and spacetime must be put in by hand in a nonperturbative
formulation and which ones may be expected to emerge dynamically. 
Since we are not likely to get something for nothing, it is a good idea to critically examine why and by what mechanism 
certain classical features should emerge from a chosen set of microscopic ingredients. 
The four-dimensionality of spacetime is a nontrivial requirement (see Q28),
but there are of course many other properties of spacetime one would like to reproduce. 
It is an appealing thought that extended spacetime may simply emerge from some ensemble of structureless quantum ``bits'' 
that have absolutely no resemblance with the degrees of freedom of general relativity, but experience with statistical
models of this type teach us that more conservative choices of ingredients are more likely to bring this about.

\section*{Q15: What were the main ideas in quantum gravity research before the year 2000?}
\label{sq15}
\addcontentsline{toc}{section}{Q15: What were the main ideas in quantum gravity research before the year 2000?}

To create a background and context for today's lines of research into nonperturbative quantum gravity, we will start
by sketching how the subject developed in the years from the early 1980s onwards.  
This important period not only saw the confirmation of the perturbative nonrenormalizability of gravity -- raising the question
of how to proceed further -- but also the birth of two major new approaches, superstring theory \cite{Green1984,Gross1984} and
loop quantum gravity \cite{Rovelli1989}. This led to a surge in activities in the respective parts of the theory community (cf.\ Q10), 
and considerable optimism that a resolution of quantum gravity was on the horizon. Although the two approaches
developed independently and use very different set-ups, covariant vs.\ canonical and unified vs.\ non-unified, 
it is rather striking that both of their dynamics involves extended objects in the form of one-dimensional strings or loops. 
An excellent review of the state of the art of quantum gravity in 2001 is \cite{Carlip2001}, with a level and scope somewhat
similar to the present lecture notes. 

\begin{figure}[t]
\centerline{\scalebox{0.48}{\rotatebox{0}{\includegraphics[angle  =270]{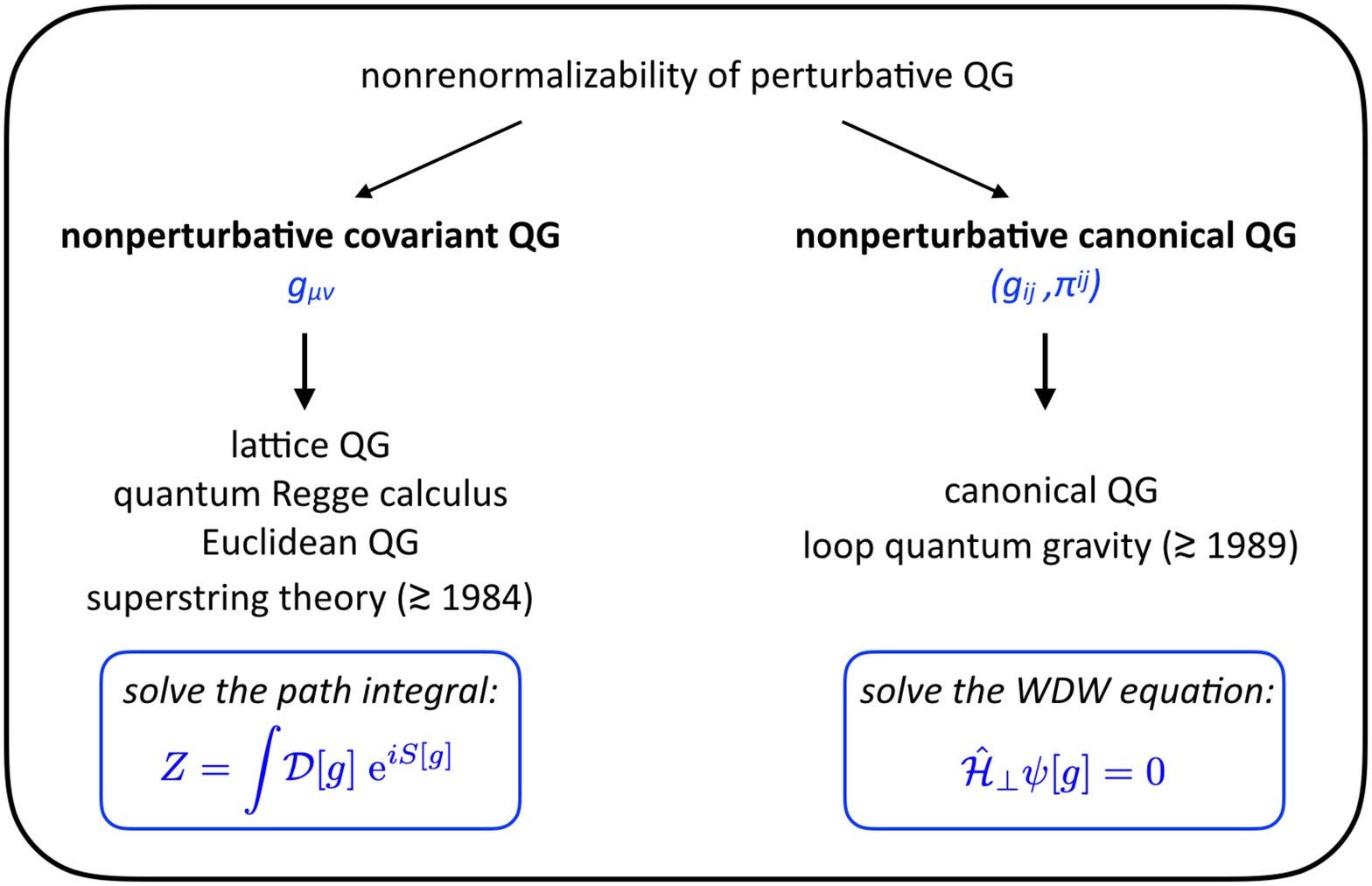}}}}
\caption{Major directions in nonperturbative quantum gravity (QG) research, ca.\ 1980--2000.}
\label{history1}
\end{figure}

A diverse research landscape in quantum gravity existed prior to the advent of strings and loops, and continued to develop
alongside them, including nonperturbative
approaches that were important milestones for later developments (see Fig.\ \ref{history1}).\footnote{The
development of quantum gravity prior to 1970 and the origin of the schism between the particle physicists'
and the general relativists' view of the subject are sketched in \cite{Isham1995}. An excellent account of the
status quo of the field in the mid-1980s is \cite{Isham1986}.} 
Most notable are lattice gravity approaches \cite{Smolin1978,Menotti1985}, inspired
by the successes of lattice quantum chromodynamics (see Q25), and quantum Regge calculus \cite{Rocek1981,Hamber1985}, 
a purely geometric formulation based on simplicial manifolds (see Q26). Canonical quantum gravity in a metric formulation,
based on the canonically conjugate field variables $(g_{ij},\pi^{ij})$ had been around for a long time \cite{DeWitt1967a,Isham1984} 
before the gauge-theoretic ``Ashtekar variables'' came along \cite{Ashtekar1986}, which provided the starting point for
an alternative canonical quantization based on the holonomy variables of loop quantum gravity (see Q21). 
On the covariant side, in so-called Euclidean quantum gravity, a
Euclidean version of the gravitational path integral had been studied since the late 1970s \cite{Hawking1979}.
In it, the counterpart $Z_{\text{eu}}$ of the functional integral (\ref{pathint}) 
is real and defined in terms of positive definite, Riemannian metrics $g_{\text{eu}}\in \text{Riem}(M)$,
\begin{equation}
Z_{\text{eu}}=\int\limits_{ \frac{\text{Riem}(M)}{\text{Diff}(M)}   }\!\!\! {\cal D} [g_{\text{eu}}]\, {\rm e}^{-S^{\rm EH}[g_{\text{eu}}]},
\label{eupathint}
\end{equation}
instead of Lorentzian metrics. Since it is difficult to make sense of the full Euclidean path integral in a continuum formulation
(cf.\ Q24), most applications have been in quantum cosmology or in a semiclassical context \cite{Gibbons1993}.

Despite the breadth and variety of these approaches, and to the extent they genuinely go beyond weak-field perturbation theory,
they tend to share some recurring challenges, most importantly, how to implement a regularization and renormalization compatible with 
diffeomorphism invariance, and how to develop effective theoretical and computational tools to get a handle on the
nonperturbative quantum dynamics.

\section*{Q16: Which technical tools are needed to make progress in quantum gravity?}
\label{sq16}
\addcontentsline{toc}{section}{Q16: Which technical tools are needed to make progress in quantum gravity?}

Having made the case for the need to go beyond perturbation theory in quantum gravity, 
let us review the technical tools at our disposal to pursue this goal. 
The methodology we can bring to bear on the problem will also have consequences for what we mean by {\it solving}
quantum gravity and what we can realistically expect to be able to achieve in the short and medium term. 
Within today's frame of reference in theoretical high-energy physics, set by the extent to which we understand the other
fundamental interactions, the magnitude of the challenge is formidable.
We want to find the quantum laws of a four-dimensional interacting
quantum theory of gravity, which at the classical level has a highly nontrivial field content and nonstandard local symmetry structure.
In addition, spacetime is not part of a fixed background structure, but should be 
determined dynamically, requiring at the very least a generalization of the standard set-up of relativistic quantum field theory to
allow for spacetime geometry to be dynamical. 

Outside perturbation theory, one usually cannot solve local, interacting quantum field theories (or underlying statistical models)
analytically. The few known examples of exactly soluble models are almost all defined in two dimensions  
and therefore not of direct physical relevance. They include statistical lattice models like the two-dimensional Ising model \cite{Guttmann2005}, 
conformal field theories \cite{Teschner2017} and several quantum field theories
in $(1+1)$ dimensions \cite{Summers2012}. An array of analytical techniques, e.g.\  from algebraic quantum 
field theory and the theory of integrable systems, has been used successfully in two dimensions, but does not generalize easily to higher dimensions. 
In view of how little we understand in a mathematically rigorous way about the nonperturbative properties of nonabelian quantum gauge field theories on flat spacetime and about general quantum field theories on fixed, curved spacetimes, expecting any kind of exact solution of quantum
gravity in the near future seems wildly unrealistic. This conclusion may seem straightforward, but is not always 
appreciated by theoreticians, since we are used to relying on the powerful tools of tensor calculus on classical, differentiable 
manifolds\footnote{Riemann, who laid the foundations for (pseudo-)Riemannian geometry in terms of a metric $g_{\mu\nu}(x)$,
already anticipated that this way of modelling space(time) might have to be revised in light of an improved microscopic physical understanding in the future \cite{Riemann1854}.}  
and to some extent may be spoiled by the 
beauty of the (few) exact solutions of the Einstein equations that dominate textbooks on general relativity.   

Given the unsuitability of perturbation theory and the lack of exact methods, 
what is required primarily are suitable approximation techniques to explore the nonperturbative regime of quantum gravity,
in whichever formulation, including tools from numerical analysis.  
The situation is similar to that in other branches of physics, like quantum many-body or statistical systems, 
where approximation methods other than perturbation theory around an exact solution are essential.
Of course, we may still discover new fundamental physical principles to complement and strengthen this
main route of exploration, although it is difficult to imagine how they could completely substitute a direct, local analysis of
strongly coupled quantum-gravitational systems.\footnote{We may still find a magic wand that gives us
indirect access to Planckian physics, in the spirit of the gauge/gravity correspondence \cite{Ammon2015}, 
but involving the strong- rather than the weak-gravity regime.} 

An important example in high-energy physics of the tools we have in mind is the use of lattice methods in quantum chromodynamics
(QCD), where one performs a first-principles evaluation of the (Wick-rotated) path integral on a spacetime 
lattice via Monte Carlo sampling \cite{Rothe2012}.  
It has led over time to spectacular advances in our quantitative understanding of many nonperturbative 
properties of the theory, which cannot be obtained by other means, including the spectrum and structure of hadrons. 
If in quantum gravity we could draw on a computational framework of similar power, sophistication and accuracy as
afforded by current QCD lattice simulations \cite{Durr2008}, we would be in an excellent place for achieving major breakthroughs. 
Namely, even before relating directly to true quantum gravity phenomenology, 
the availability of nonperturbative computational tools serves several purposes: 
(i) checking whether a given candidate theory is sufficiently well-defined and complete
(vis-\`a-vis the ingredients specified in Q13) to allow computations to be set up in the first place, 
without introducing additional ad-hoc assumptions
and free parameters, (ii) establishing its physical properties, including the classical limit, by computing suitable quantum observables 
(cf.\ Q29) and comparing its results with those of other approaches, (iii) using the outcome of ``numerical experiments'' to check conjectures and hypotheses, and provide feedback for further theory-building.  

The development of efficient numerical approximation methods suited to modelling the dynamics of four-dimensional
quantum gravity is nontrivial, and
has to overcome both conceptual and technical difficulties, that is, {\it what} to compute and {\it how}.
With regard to either of these aspects, two main types of methodology have been developed and improved continuously over the last 
couple of decades, and are furthest along in producing results. Firstly, one can use
Markov chain Monte Carlo methods \cite{Newman1999,Hanada2018} to approximate gravitational path integrals, usually regularized 
in terms of simplicial geometries, as is the case in the approaches of (causal) dynamical triangulations (see Q27 and \cite{Ambjorn2012,Loll2019}) 
and quantum Regge calculus (see Q26 and \cite{Loll1998,Williams2009}). Secondly, there are
functional renormalization group methods \cite{Gies2006}, which are centred on solving a 
(truncated) flow equation for a scale-dependent effective action, using some approximation scheme. 
They are applied to quantum gravity in the approach of asymptotic safety (see Q23 and \cite{Reuter2019}).

In the quantum gravity community, which is traditionally mathematics-oriented and focused on exact methods, 
there is a growing recognition that the development of nonperturbative computational methods 
must be part of any research strategy whose aim is 
a quantum gravity theory that can produce numbers, i.e.\ quantitative predictions that can ultimately be compared with nature. 
The methodologies mentioned above have already had a significant impact on the field, since they have been used to 
identify and compute specific quantum observables (cf.\ Q28, Q29). In the process, they have also opened up new perspectives 
on the possible nature of quantum gravity at the Planck scale and on what we can expect to be able to compute.

\section*{Q17: Can we learn from simplified models about quantum gravity proper?}
\label{sq17}
\addcontentsline{toc}{section}{Q17: Can we learn from simplified models about quantum gravity proper?}

Perhaps unsurprisingly, the {\it specific} lessons that can be learned for full, four-dimensional quantum gravity 
by studying simplified models are rather limited. 
What usually makes such models more tractable -- although by no means trivial -- is the absence of 
precisely those features that are related 
to key problems of the full theory, namely, the presence of propagating degrees of freedom and the ensuing need to renormalize in a way
that respects gravity's background independence. Nevertheless, simplified models can sometimes be useful as 
a test bed for new ideas or calculational
techniques, or to illustrate some general properties of diffeomorphism-invariant quantum systems. 
We will describe briefly two different sources of simplification, spacetimes with fewer than four dimensions and 
spacetime geometries with global symmetries. 

{\it Dimensionally reduced models of quantum gravity} are very popular, especially in dimension 2, where various solution
techniques are available (including combinatorial, matrix model and transfer matrix methods \cite{David1993,DiFrancesco1993,Ambjorn1999}),
and some models can be solved exactly. Gravity in $D$ spacetime dimensions is usually defined by a 
$D$-dimensional version $S^{(D)}$ of the Einstein-Hilbert action (\ref{SEH}),
\begin{equation}
S^{(D)}[g]=\frac{1}{16\pi G_{\rm N}}\, \int_{M} d^D x\, \sqrt{|\det(g)|}\, (R -2 \Lambda ),
\label{daction}
\end{equation}
on a $D$-dimensional manifold ${M}$ with Lorentzian metric $g_{\mu\nu}$. Although the functional form of the Lagrangian 
in terms of the metric looks identical for all $D$, the corresponding classical theories are very different and have 
$D(D-3)/2$ local physical degrees of freedom for $D\! >\! 2$ (and none for $D\! \leq\! 2$). In particular, unlike in $D\! =\! 4$, there are no local
propagating field excitations in $D\! =\! 2,3$. 

In $D\! =\! 2$ there are not even classical equations of motion (and therefore no classical
solutions), since the integral over the Ricci scalar $R$ is a topological invariant. The path integral for pure gravity in two dimensions can be solved analytically, both by continuum methods and as the continuum limit of statistical models of random geometry. The possible outcomes
fall into just two universality classes, characterized by certain critical exponents. The universality class of 2D Euclidean quantum gravity 
is characterized by a Hausdorff dimension $d_H\! =\! 4$, spectral dimension $d_S\! =\! 2$ and entropy exponent $\gamma\! =\! -1/2$,
and contains Liouville quantum gravity \cite{David1988,Distler1988}, 2D Euclidean dynamical triangulations \cite{David1984,Kazakov1985} 
and the Brownian sphere \cite{LeGall2014}. The
other universality class is that of Lorentzian or causal models, characterized by $d_H\! =\! 2$, $d_S\! =\! 2$ and $\gamma\! =\! 1/2$ and
contains 2D gravity in proper-time gauge \cite{Nakayama1993}, 2D causal dynamical triangulations \cite{Ambjorn1998} and the uniform infinite planar tree \cite{Durhuus2009}. 
Because of the absence of a classical limit, these models are ``purely quantum'' and in this sense 
have little to do with quantum gravity proper. Nevertheless, they are mathematically rich and 
can be coupled to dilatonic and other (conformal) matter in many variants (see \cite{Grumiller2002,Mertens2020} and references therein).  

In $D\! =\! 3$, the classical solutions of pure gravity are exceedingly simple: for vanishing cosmological constant, spacetime is flat everywhere.
The only dynamical degrees of freedom are finitely many global modes of the metric \cite{Moncrief1989}, which 
are only present when the spacetime $M$ has a nontrivial topology. 
However, this does not lead to any obvious simplifications when one tries to solve the path integral for the three-metric $g_{\mu\nu}(x)$, compared with the four-dimensional case.
What seems to make the quantum theory of 3D gravity more tractable after all is the
existence of an alternative, gauge-theoretic Chern-Simons formulation of the three-dimensional Poincar\'e group \cite{Witten1988}, 
which allows for an explicit reduction to the physical, finite-dimensional covariant phase space. 
However, this is not completely satisfactory from a physics point of view:
since both time and spacetime geometry have disappeared from the formulation, it is difficult to address any questions about the spacetime 
dynamics in the corresponding quantum theory. In addition, the presence of degenerate dreibeins (corresponding to metrics
with $\det g\! =\! 0$, cf.\ \cite{Matschull1999}) may well lead to a quantum theory that is different from another 
dozen or so quantum theories of 3D gravity,
covariant and canonical, first- and second-order, Lorentzian and Euclidean (see \cite{Loll1995,Carlip1998,Carlip2004} and references therein). 
Unfortunately, beyond the simple case of an empty universe with toroidal spatial slices, $M\! =\! \mathbb{R}\times T^2$, little is known about the equivalence or otherwise of these approaches. This is due to their incompleteness, many technical complications and a 
general lack of quantum observables, whose spectra one might otherwise compare. 
Of particular interest is the question of whether there are well-defined quantum operators associated with length and area
measurements, and whether their spectra are discrete or continuous \cite{Freidel2002,Budd2009,BenAchour2013,Wieland2017}.

A different type of simplification occurs in {\it symmetry-reduced models of quantum gravity}. In these models, 
isometries like spatial homogeneity or spherical symmetry are imposed on spacetimes,
leading to a configuration space that is much smaller than $\text{Lor}(M)$.
Classically, one distinguishes between minisuperspace models, which have finitely many degrees of freedom, and
midisuperspace models, which have field-theoretic degrees of freedom. 
For example, in the case of the homogeneous and isotropic FLRW metrics used in standard cosmology, the configuration space 
of local metric fields $g_{\mu\nu}(x)$ is reduced to a single, time-dependent parameter, the scale factor $a(t)$.
The quantization of this and other spatially homogeneous models is the subject of quantum cosmology \cite{Wiltshire1995,Kiefer2012,Ashtekar2011,Bojowald2020}. Its main motivation is the hope that these 
drastically simplified systems can capture key aspects of the
quantum behaviour of the universe near the big bang, e.g.\ by providing models for singularity resolution. 
Other symmetry-reduced models whose quantization has been studied are spherically symmetric gravitational systems\footnote{These should not be confused with semiclassical treatments of black holes, where the gravitational degrees of freedom are not quantized.}, 
Einstein-Rosen cylindrical waves and Gowdy models (see \cite{BarberoG2010} for an authoritative review). 
Identifying and counting the physical degrees of freedom of these systems tends to be subtle, 
and dealing with their quantum dynamics even more so. Challenges and ambiguities exist with regard to 
the implementation of the reduction,
the treatment of (residual) diffeomorphisms, factor-ordering issues and the potential noncommutativity of quantization and symmetry reduction. There are no known mechanisms to lift the results and solutions of symmetry-reduced 
quantum systems to full quantum gravity.

\section*{Q18: Are covariant and canonical quantum gravity equivalent?}
\label{sq18}
\addcontentsline{toc}{section}{Q18: Are covariant and canonical quantum gravity equivalent?}

One of the crossroads highlighted in Q13 was the choice between a covariant and a canonical ansatz for the nonperturbative
quantum dynamics. In the context of quantum mechanics and relativistic quantum field theory, we are used to thinking about these
two approaches as complementary, but ultimately equivalent. When it comes to nonperturbative quantum gravity, the incompleteness of 
existing formulations and their often very different set-ups make a detailed assessment of their equivalence difficult. 
Nevertheless, since quantum gravity is a theory of dynamical spacetime, there may be reasons why
either a path-integral or an operator-algebraic approach is preferable or more suited to the problem {\it in principle}.

Let us recall the main features of the two formulations for the case of quantum field theories on Minkowski space.
The starting point of the covariant (``spacetime'') formulation is the classical Lagrangian density $\mathcal L[\phi]$ of some
fields $\phi$, whose associated action $S\! =\! \int\! d^4x\, {\mathcal L}$
appears in the amplitudes $\exp (iS)$ of the path integral (\ref{pathint}). The latter serves as a generating functional of 
$n$-point functions, and Poincar\'e covariance is manifest. 
By contrast, the canonical (``space+time'') formulation requires an explicit choice of time, with respect to which the field momenta 
$\pi(x)\! :=\! \delta {\mathcal L}/\delta \dot{\phi}(x)$
and a Hamiltonian are defined. The quantization proceeds via the equal-time canonical commutation relations 
\begin{equation}
[\hat{\phi} (x),\hat{\pi} (y)]|_{x^0=y^0}=i \delta^{(3)}(\vec{x}-\vec{y})
\label{ccr}
\end{equation}
of the elementary field operators,
and an analysis of the spectrum of the self-adjoint Hamiltonian operator $\hat H$. Unitarity, in the form of the unitary
time evolution operator $\hat{U}(t)\! =\! \exp (-i\hat{H} t)$, is manifest. In perturbative computations, especially in
gauge field theory, the covariant framework is much preferred because of its explicit Poincar\'e covariance
and systematic treatment of the gauge degrees of freedom, but this is an issue of convenience rather than principle. 

Quantum gravity beyond perturbation theory on flat space is neither Poincar\'e-covariant nor subject to
local gauge transformations. Instead, it should be invariant under a suitable implementation of four-dimensional diffeomorphisms. 
If a covariant continuum framework is chosen, this will involve either a unitary representation of the infinite-dimensional Lie
group $\text{Diff}(M)$ or, more commonly, a suitable gauge fixing. However, things look different in a canonical continuum description,
where after a (3+1)-decomposition a more complicated, projected version of the algebra of diffeomorphism generators appears 
(cf.\ Q19), whose representation in the quantum theory turns out to be a major source of difficulty. This could be an indication that 
canonical quantization is doomed, but -- as happens frequently in quantum gravity -- it is not easy to distinguish between conceptual and
``mere'' technical problems (that may still be overcome), let alone prove no-go theorems. 
Alternative covariant quantum approaches based on simplicial 
geometries (see Q26, Q27) evade the issue of representing the diffeomorphisms altogether.
This methodology does not have an obvious canonical analogue. 

A potential argument in favour of canonical methods is that nontrivial algebras of
operators, replacing the flat-space commutators (\ref{ccr}), may be able to capture the nonlinearities of gravity and the space of geometries. 
This is illustrated by the example of quantum mechanics on $\mathbb{R}_+$, the positive real line, for which the usual Schr\"odinger
quantization with commutators $[\hat{x},\hat{p}]\! =\!i$ is inadequate, since the spectrum of the self-adjoint operator $\hat x$
is the entire real line, and not the desired half-line $\mathbb{R}_+$. 
A possible resolution is to promote the classical phase space functions $x$ and $xp$ to self-adjoint quantum operators, instead of $x$ and $p$. 
They obey the affine commutation relation $[\hat{x},\hat{xp} ]\! =\! i \hat{x}$, which does allow for irreducible representations with the
correct spectrum $\text{spec}(\hat{x})\! =\! \mathbb{R}_+$. It has been argued that the condition $\det g_{ij}\! >\! 0$ in canonical
gravity should be incorporated in an analogous fashion \cite{Klauder1970,Isham1983}. A similar philosophy underlies the framework
of algebraic quantum field theory on curved spacetime \cite{Fewster2015}, but there
it is unclear how the key property of {\it Einstein causality}, 
the commutativity of observables in spacelike separated regions, can be generalized to a situation where the metric (and therefore
the causal structure) is dynamical \cite{Giddings2015}.

Once we have found a viable quantum theory of gravity, it may well have complementary covariant and canonical formulations, but until that
time the pendulum may still swing back and forth, as it has done in the past. 
The momentum is currently firmly on the covariant side (cf.\ Q22), despite a couple of reputational setbacks since the 1970s, related to the
nonrenormalizability of the perturbative path integral and the ill-defined nature of the path integral in Euclidean quantum gravity (cf.\ Q24).
Partly due to the development of new, nonperturbative 
methods, path integrals seem to be undergoing a renaissance. 
Noteworthy in this context is the advent of a new rigorous mathematical
framework to treat {\it quantum geometry} or {\it random geometry} in two dimensions \cite{Sheffield2022}.\footnote{This grew 
out of already mentioned work from the 1980s \cite{David1984,DiFrancesco1993}, which tried to understand
the nonperturbative dynamics of string worldsheets \cite{Ambjorn1985}, in what today is called noncritical string theory.} 
It provides an encouraging proof of principle for the methodology of obtaining a nonperturbative theory of quantum gravity as
a mathematically well-defined scaling limit of a sum over discrete random geometries, albeit for a two-dimensional toy model.
Efforts are under way to generalize elements of the mathematical framework to higher dimensions \cite{Budd2022}.
Ahead of any such developments, physicists have long ago embarked on applying the same construction principle to 
nonperturbative quantum gravity in Lorentzian signature in four dimensions, with considerable success (cf.\ Q27). 
To what extent a theory of quantum gravity can ultimately be shown to exist and
described ``in closed form'' in terms of suitable mathematical structures is an interesting question, 
even though no-one has yet set out any prize money for answering it, unlike
for the corresponding question in nonabelian gauge theory \cite{Clay}.

\section*{Q19: What is the starting point for a canonical quantization of gravity?}
\label{sq19}
\addcontentsline{toc}{section}{Q19: What is the starting point for a canonical quantization of gravity?}

Going from a covariant to a canonical description requires the choice of a global time. In general relativity this
is always possible since one usually considers spacetimes $(M,g_{\mu\nu})$ which are globally hyperbolic. This implies the
existence of a foliation of $M$ into spacelike Cauchy hypersurfaces and that $M$ has the form of a topological product $M\! =\! \mathbb{R}\times\Sigma$ of a time direction and a three-dimensional manifold $\Sigma$ \cite{Minguzzi2019}. The standard way to
perform the (3+1)-decomposition of the four-metric and the Einstein equations is the so-called ADM formalism \cite{Misner1973,Giulini2015}, which was developed around 1960.\footnote{see Sec.\ 4 of \cite{Blohmann2010} for a brief history of the canonical formulation} 
The ADM reference foliation $\{ \Sigma_t, \, t\!\in\!\mathbb{R} \}$ should be viewed as a background
structure needed to obtain a Hamiltonian dynamical system. 
Note that the real parameter $t$ labelling 
the leaves of the foliation is merely a ``coordinate time'' and does not represent a physically distinguished notion of time. 
The Lorentzian metric $g_{\mu\nu}$ is projected along and perpendicular to the foliation, leading to a Hamiltonian picture with
conjugate variable pairs $(g_{ij},\pi^{ij})$, satisfying the nonvanishing, elementary Poisson bracket relations
\begin{equation}
\{g_{ij}(\vec{x},t),\pi^{kl}(\vec{y},t)\}=\frac{1}{2} (\delta_i^k\delta_j^l+\delta_i^l \delta_j^k )\, \delta^{(3)}(\vec{x},\vec{y}).
\label{poissongpi}
\end{equation}
As a consequence of diffeomorphism invariance, one obtains a Hamiltonian constrained system with a set of four first-class
constraints (in Dirac's classification \cite{Dirac1964,Sundermeyer1982}), namely,
\begin{equation}
\mathcal{H}_i[g,\pi]=0,\hspace{1cm}\mathcal{H}_\perp [g,\pi] =0,
\label{constraints}
\end{equation}
whose precise functional form is not important for our present discussion.
Physical configurations lie on the constraint surface $\mathcal C$, the subspace of the phase space $T^*\text{Riem}(\Sigma)$ defined by (\ref{constraints}). As usual for a first-class constrained system, two configurations on $\mathcal C$ are physically equivalent if they differ by a canonical transformation whose infinitesimal generators are the phase space functions $\mathcal{H}_i$ and $\mathcal{H}_\perp$.
The three constraint functions $\mathcal{H}_i$ generate spatial diffeomorphisms along the spatial manifold 
$\Sigma$, and $\mathcal{H}_\perp$
generates deformations of $\Sigma$ normal to itself in $M$ (when the equations of motion are satisfied).
Together they form a closed algebra under Poisson brackets \cite{Katz1962},
\begin{subequations}\label{katz}
\begin{align}
    \left\{\mathcal{H}_i(x),\mathcal{H}_j(y)\right\}=\; &\mathcal{H}_i (y)\partial^x_j\delta^{(3)} (\vec{x},\vec{y})-\mathcal{H}_j (x)\partial^y_i\delta^{(3)}(\vec{x},\vec{y} )\label{katz1}\\
    \left\{\mathcal{H}_i(x),\mathcal{H}_\perp(y)\right\}=\; &\mathcal{H}_\perp (x) \partial^x_i\delta^{(3)} (\vec{x},\vec{y} )\label{katz2}\\
    \left\{\mathcal{H}_\perp(x),\mathcal{H}_\perp(y)\right\}=\; &g^{ij}(x)\mathcal{H}_i (x)\partial^y_j\delta^{(3)} (\vec{x},\vec{y} )-
    g^{ij}(y)\mathcal{H}_i (y) \partial^x_j\delta^{(3)} (\vec{x},\vec{y} ).\label{katz3}
\end{align}
\end{subequations}
Noteworthy here is that (\ref{katz1}) are the Lie algebra relations of the spatial diffeomorphism group $\text{Diff}(\Sigma)$, 
and -- perhaps surprisingly -- that (\ref{katz1})-(\ref{katz3}) do {\it not} form a Lie algebra. 
The reason for the latter is the appearance of field-dependent
structure functions (depending on the inverse metrics $g^{ij}$) rather than structure constants on the right-hand side of (\ref{katz3}). 
Because the introduction of a foliation breaks the manifest diffeomorphism symmetry of spacetime, the constraint algebra is
not the Lie algebra of the diffeomorphism group $\text{Diff}(M)$, but a modified version projected along and normal to
the hypersurfaces $\Sigma$. The exact functional form of the algebra relations captures the four-dimensional diffeomorphism
invariance at the canonical level. This is relevant in the quantum theory if one proceeds \`a la Dirac by first quantizing and then
implementing the constraints: only the presence of an anomaly-free quantum version of the algebra (\ref{katz}) 
ensures diffeomorphism invariance and the possibility to recover spacetime properties from a canonical formulation.
Of equal importance are the constraints (\ref{constraints}), because it can be shown that if they hold for arbitrary foliations, the
Einstein equations are satisfied too \cite{Isham1992}. Another key element of the canonical quantization is therefore the imposition of
quantum analogues of the equations (\ref{constraints}).

\section*{Q20: What is the Wheeler-DeWitt equation?}
\label{sq20}
\addcontentsline{toc}{section}{Q20: What is the Wheeler-DeWitt equation?}

The Wheeler-DeWitt equation is the central dynamical equation of canonical quantum gravity. We will discuss it first in the context of 
quantum geometrodynamics, the 
quantization of the metric formulation \`a la ADM introduced in Q19, where the phase space variables $(g_{ij},\pi^{ij})$
are turned into elementary operators $(\hat{g}_{ij},\hat{\pi}^{ij})$. The latter are taken to satisfy the nonvanishing 
canonical commutation relations
\begin{equation}
[ \hat{g}_{ij}(\vec{x},t),\hat{\pi}^{kl}(\vec{y},t) ]=\frac{i}{2} (\delta_i^k\delta_j^l+\delta_i^l \delta_j^k )\, \delta^{(3)}(\vec{x},\vec{y}),
\label{hatgpi}
\end{equation}
which are the quantum analogues of the Poisson brackets (\ref{poissongpi}). 
In a formal quantization mimicking the Schr\"odinger representation of quantum mechanics, the operators $\hat{g}_{ij}$ are
represented  on wave functionals
$\psi [g_{ij}]$ by multiplication and the $\hat{\pi}^{ij}$ by ($-i$ times the) functional differential with respect to $g_{ij}$. Following the logic of the Dirac quantization of constrained systems, quantum solutions or physical
wave functionals $\psi^{\text{ph}}[g_{ij}]$ are those that are ``annihilated'' by operator versions of the constraints (\ref{constraints}),
\begin{equation}
    \hat{\mathcal{H}}_i [\hat{g},\hat{\pi}] \psi^{\text{ph}}[g]=0,\;\;\;\;\;
    \hat{\mathcal H}_\perp [\hat{g},\hat{\pi}]\psi^{\text{ph}}[g]=0.
\label{qconst}
\end{equation}
The equations involving the quantized momentum constraints $ \hat{\mathcal{H}}_i$ are formally easy to solve, by selecting
states $\psi$ that depend only on equivalence classes $[g_{ij}]$ of metrics under the action of spatial diffeomorphisms. This is not
the case for the equation involving the quantized Hamiltonian constraint $\hat{\mathcal H}_\perp$, which is the Wheeler-DeWitt equation we already encountered in eq.\ (\ref{wdw}) above.

The program of quantum geometrodynamics as an ansatz for tackling 
the full theory fizzled out during the 1980s, because of the ill-defined mathematical nature of the set-up and multiple problems regarding the Wheeler-DeWitt equation \cite{Kuchar1993,Kiefer2012}.  
Note that to obtain the explicit functional form (\ref{wdw}), a specific operator ordering was chosen (with all the $\hat{\pi}^{ij}$ to the right). 
However, the associated operator-ordering ambiguities cannot be addressed and resolved without at the same time
regularizing the Hamiltonian operator \cite{Tsamis1987,Friedman1988}. The singular product of the two functional
derivatives in (\ref{wdw}) must be regularized, but it is unclear how to do this while maintaining diffeomorphism invariance. 
To illustrate the issue, a standard point-splitting regularization will introduce a dependence on the metric, which however is a dynamical 
variable and will therefore affect the quantum version of the constraint algebra (\ref{katz}).

Even if one could find solutions to the properly regularized and renormalized Wheeler-DeWitt equation, there would still be nontrivial 
work to do because of the infamous ``problem of time'' \cite{Isham1992,Kiefer2012}. Since the equation does not refer to any notion of
time, time must be recovered somehow from the physical states $\psi^{\text{ph}}$. Even more challenging, (quantum) spacetime
must also be recovered, together with a classical limit, to make sure that the theory describes gravity.\footnote{Note that in
these lecture notes we are treating quantum gravity mainly in a localized sense, without considering any interpretational, 
quantum-cosmological issues that arise when 
the whole universe is regarded as a quantum system, and where $\psi^{\text{ph}}$ may describe a ``wave function of the universe''.}
It is noteworthy that DeWitt, one of the scientists the equation is named after, 
did not think highly of it \cite{DeWitt1997}: ``\emph{I think it is a bad equation, for the following 
reasons: (1) By focussing on time slices (spacelike 3-geometries) it violates the
very spirit of relativity. (2) Scores of man-years have been wasted by researchers trying to extract from it a natural time parameter. (3)
Since good path-integral techniques exist for basing quantum theory on gauge invariant observables only, it seems a pity to drag
in the paraphernalia of constrained Hamiltonian systems.}'' Of course, these points of criticism are by no means no-go theorems, 
and new insights may lead to unexpected breakthroughs in dealing with the Wheeler-DeWitt equation.

\section*{Q21: What is loop quantum gravity?}
\label{sq21}
\addcontentsline{toc}{section}{Q21: What is loop quantum gravity?}

Loop quantum gravity (LQG) is a variant of canonical quantum gravity which is based on an alternative parametrization of canonical
general relativity in terms of conjugate variable pairs $(A_i^a,E^i_a)$ instead of the ADM variables $(g_{ij},\pi^{ij})$ 
\cite{Rovelli2008,Thiemann2007}. Up to a density weight, $E^i_a(x)$ is a dreibein (triad), where the
indices $a,b$ refer to an internal vector space with Euclidean metric $\delta_{ab}$, and $A_i^a(x)\in \cal{A}$ is a 
$SO(3)$-connection of Yang-Mills type. In addition to being diffeomorphism-invariant, which is associated with analogues of
the four constraints (\ref{constraints}), this formulation is also invariant under local $SO(3)$-frame rotations, implying three additional
Gauss constraints $G_a [A,E]\! =\! 0$. The phase space of gravity therefore appears as a subspace of
the phase space $T^*{\mathcal A}$ of a Yang-Mills theory.

This new ``connection dynamics'' offers a fresh, gauge-theoretic perspective on canonical
quantum gravity. At the same time, it is complementary to the old geometrodynamics, in the sense that in the pair $(A,E)$, metric and
momentum have swapped roles, since $E$ depends on the spatial metric $g$ and $A$ on its momentum $\pi$ 
(in essence, the extrinsic curvature). 
It suggests a quantization where $\hat A$ acts by multiplication and $\hat E$ by functional differentiation with respect to
$A$, which may lead to an inequivalent quantum theory, reminiscent of the Chern-Simons formulation
of three-dimensional quantum gravity (cf.\ Q17). 

A crucial ingredient borrowed from Yang-Mills theory are holonomies. Recall that
the holonomy $U_\gamma[A]$ associated with a path $\gamma\!\subset\!\Sigma$ and a gauge connection $A$ is given by the 
path-ordered exponential of $A$ along $\gamma$,\footnote{If the path is closed, a gauge-invariant Wilson loop functional $W_\gamma[A]\! =\!\text{Tr}\, U_\gamma$ is obtained by taking the trace. Wilson loops were used in the
original construction of loop quantum gravity \cite{Rovelli1989}, explaining the origin of the word ``loop''.}
 \begin{equation}
U_\gamma[A]={\text P} \exp \int_\gamma A.
\label{holo}
\end{equation}
The starting point of LQG is a classical holonomy-flux (Poisson) algebra,
where the holonomies are associated with one-dimensional edges of a graph embedded in $\Sigma$
and the fluxes are obtained by integrating the (two-form dual to the) field $E$ over two-dimensional surfaces 
in $\Sigma$ \cite{Ashtekar2004,Sahlmann2010}. 
{\it The} key assumption of LQG, from which (almost) everything else follows, is that these holonomies and fluxes 
are represented by well-defined operators in the quantum theory. The assumption is not trivial, since these quantities are
peaked in a highly discontinuous manner on paths and surfaces in the spatial manifold $\Sigma$, and it is not clear that this is
a physically appropriate choice. For example, Wilson loops in QCD 
must be regularized and renormalized, and past attempts to use them as fundamental building blocks of the
theory have not been successful \cite{Loll1992}. For gravity, it has been argued that background independence is a game changer, but only the outcome of a viable quantum gravity theory can ultimately validate LQG's main conjecture.  

What then follows \cite{Lewandowski2005} is a mathematically well-defined construction of LQG at the kinematical level, i.e.\ prior to 
solving the Wheeler-DeWitt equation. A basis for the kinematical Hilbert space is given by so-called spin network states,
oriented graphs embedded in $\Sigma$ with certain edge and vertex labels, which can be thought of as distributional excitations of
the connection along the graph. The special nature of these states allows for an implementation of both gauge and spatial diffeomorphism invariance. Remarkably, even a concrete proposal for how to represent the Hamiltonian constraint operator $\hat{\mathcal H}_\perp$ 
was made in this setting \cite{Thiemann1996}. 

Compared to quantum geometrodynamics, this looks like good progress,
but significant challenges remain \cite{Nicolai2005,Thiemann2006}. The proposed operator $\hat{\mathcal H}_\perp$ 
is highly ambiguous due to ordering issues, and may not
be physically correct: the quantum version of the constraint algebra (\ref{katz}) closes, but the operator version of the field-dependent 
coefficients on the right-hand side of (\ref{katz3}) comes out wrong \cite{Thiemann2020}. Work continues on modifications of
the kinematical framework that may have a more immediate physical interpretation (see e.g.\ \cite{Bahr2015}).
Recent studies of the Hamiltonian operator in simplified models indicate 
that interesting solutions to the Wheeler-DeWitt equation may generically be non-normalizable with respect to the scalar product 
of the kinematical Hilbert space \cite{Thiemann2021}. This is relevant to the longstanding issue of
whether the well-known kinematical LQG result, according to which the area and volume operators have discrete spectra, persists in the physical
Hilbert space \cite{Dittrich2007}. However, as long as there is no breakthrough on the quantum Hamiltonian constraint, on solutions to  
$\hat{\mathcal H}_\perp \psi^{\text{ph}}\! =\! 0$ and on the consistency of the quantum constraint algebra in the full theory, it will be difficult to say anything definite about the physical properties of the resulting quantum spacetime, be it on Planckian scales or in a classical limit.

\section*{Q22: How has quantum gravity research been developing since the year 2000?}
\label{sq22}
\addcontentsline{toc}{section}{Q22: How has quantum gravity research been developing since the year 2000?}

The landscape of quantum gravity has changed significantly over the last 20 years.
Almost all research conducted during this period has been in the context of covariant approaches. 
Also a large fraction of the LQG community has migrated towards covariant methods\footnote{or has focused on reduced models, 
like quantum
cosmology \cite{Banerjee2011,Bojowald2020a} or black holes \cite{Perez2017,Bojowald2020b}}, mostly to the area of so-called spin foam models \cite{Perez2012,Bianchi2017}, because of the difficulties with tackling the quantum Hamiltonian constraint directly. This array of
models is very much inspired by the ingredients of loop quantum gravity, although a precise relation 
has yet to be established \cite{Thiemann2013} and there is no general agreement on any specific model, on how to restore diffeomorphism invariance or how to implement a continuum limit.
The fact that the ``loop'' label is sometimes used to also describe these new activities
can be confusing to outsiders, because the nature of the research is rather different from that of the pre-2000s.
Something similar is true for the ``string'' label, where over roughly the same time period the bulk of research work of the (former) superstring community has not been on strings as such, indicating a shift of emphasis away from quantum gravity, 
as already mentioned in Q12.  

During the same time, work on nonperturbative approaches to quantum gravity based on quantum field theoretic methods has progressed
at an encouraging pace, 
with an ever better understanding of how background independence and dynamical geometry should be accounted for, and with
the development of efficient computational approximation methods to explore the Planckian regime in quantitative terms.
Furthest developed in terms of these aspects are a pair of complementary lattice and continuum formulations. 
Firstly, there is the approach of ``Causal Dynamical Triangulations'' \cite{Loll2007,Ambjorn2010,Ambjorn2013}, whose strategy 
is to obtain quantum gravity as the scaling limit of a regularized, geometric path integral, generalizing the corresponding exact toy model in two
dimensions (cf.\ Q17, Q18). Secondly, there is the approach of ``Asymptotic Safety'' \cite{Niedermaier2006,Reuter2012,Percacci2017}, whose aim is to solve renormalization group flow equations for a particular, 
scale-dependent effective action. These candidate theories for quantum gravity will be described in Q23 and Q27 below.  

The overall trend in nonperturbative quantum gravity appears to be away from exotic to more conservative set-ups and ingredients, 
with a renewed appreciation for the tools of quantum field theory, critical phenomena 
and the renormalization group, adapted to
dynamical spacetimes. At the same time, it should be clear from many of our earlier considerations that nonperturbative approaches which are amenable to
numerical or other approximation methods have an enormous advantage over formal, theoretical approaches which lack any such quantitative checks of their validity. With insufficient computational control, it may be very difficult or even impossible to show that a given theory 
(i) exists nonperturbatively {\it and} (ii) possesses a classical limit compatible with general relativity. Some of these models are well worth
exploring, like the causal set approach \cite{Henson2006,Surya2019,Dowker2019} or the spin foam models mentioned above, 
but we will not cover them further in this limited set of lecture notes.

\section*{Q23: What is asymptotic safety?}
\label{sq23}
\addcontentsline{toc}{section}{Q23: What is asymptotic safety?}

Asymptotically safe gravity \cite{Percacci2017,Reuter2019,Nink2013,Niedermaier2006,Niedermaier2006a} is a covariant 
approach to quantum gravity, 
which is described in the language of continuum quantum field theory, up to now mostly in a Euclidean regime\footnote{see 
Sec.\ 10 of \cite{Bonanno2020} for a discussion of Lorentzian signature in asymptotic safety and further references}. 
Following a suggestion by
Weinberg \cite{Weinberg1979}, one investigates whether quantum gravity may be {\it non}perturbatively renormalizable. 
In technical terms, this means that running coupling constants of the theory, rescaled by some common reference 
scale to make them dimensionless, should approach constant values as the energy is taken to infinity, in such a way that all
physical observables stay finite there. Equivalently, the theory is said to have an ultraviolet fixed point with
respect to the action of the renormalization group (RG), which governs the behaviour of the coupling constants as a function of
scale. If such a fixed point can be shown to exist, it implies the existence of a well-defined quantum gravity theory 
without divergences at arbitrarily high energies. For phenomenological reasons, the fixed point should be connected via an RG trajectory in coupling-constant space
to a low-energy region reproducing general relativity,

Instead of working directly with the path integral,
the key quantity one usually studies with the help of functional renormalization group methods \cite{Gies2006,Reuter1996} 
is an effective average action $\Gamma_k$, a generating functional of 
one-particle irreducible correlation functions, which depends on the metric fields, a momentum reference
scale $k$, and a regulator function $R_k$, which among other things ensures the absence of infrared and ultraviolet 
divergences in $\Gamma_k$.   
Of course, physical results should not depend on the choice of $R_k$.
One must then solve a functional differential {\it flow equation} for $\Gamma_k$. Using an ansatz 
\begin{equation}
\Gamma_k={\textstyle\sum\limits_{\alpha =1}^\infty}\, c_{\!\alpha\! }(k)\, {\mathcal O}_\alpha
\label{gammaas}
\end{equation}
in terms of the $k$-dependent coupling constants $c_{\!\alpha\! } (k)$ and a basis $\{ {\mathcal O}_\alpha \}$
of integrals of monomials in the fields and their derivatives leads to a system of nonlinear ordinary differential equations for 
the $c_{\!\alpha\! } (k)$. To preserve a form of diffeomorphism invariance, one employs the background field approximation, 
usually based on the linear split 
\begin{equation}
g_{\mu\nu}=\bar{g}_{\mu\nu}+h_{\mu\nu}
\label{background}
\end{equation}
of the ``dynamical metric'' $g_{\mu\nu}$ into a fixed background metric $\bar{g}_{\mu\nu}$ and fluctuations $h_{\mu\nu}$ \cite{Niedermaier2006,Pawlowski2020}, such that $\Gamma_k$ and the ${\mathcal O}_\alpha$ depend on both $g$ and $\bar{g}$. 
Importantly, the background metric is also used to provide the momentum scale $k$. 

To solve the flow equation, the infinite-dimensional space spanned by the operators ${\mathcal O}_\alpha$
must be truncated and the flow equations for the corresponding couplings $c_{\!\alpha\! } (k)$ be solved.
The remarkable result of the asymptotic safety research program is that for a whole array of truncations, backgrounds, gauge 
fixings and computational approaches an 
ultraviolet fixed point has been found, starting with the so-called Einstein-Hilbert truncation, which is based on just the 
Newton and cosmological constants \cite{Reuter1996,Reuter2001}.\footnote{references \cite{Bonanno2020,Pawlowski2020} 
contain compact overviews of results and \cite{Eichhorn2018} discusses applications involving matter coupling} 
This provides nontrivial evidence that quantum gravity may indeed be nonperturbatively renormalizable.
Since it is difficult to assess the validity of the truncations within the functional RG framework itself, it would be desirable
to independently corroborate the existence of fixed points in nonperturbative lattice approaches like CDT quantum gravity (cf.\ Q27). 
It is not easy to assess {\it how} nonperturbative the current pursuit of asymptotic safety is; while integrating up flow lines
in coupling-constant space clearly goes beyond perturbation theory, many of the ingredients and techniques used
are from perturbative quantum field theory. For example, one may wonder to what extent the formalism accounts for
the nonlinear structure of the space ${\mathcal G}(M)$ of geometries.

The number of free parameters of the theory, characterizing its predictive character, is equal to the 
number of linearly independent UV-attractive directions of the RG flow at the fixed point. For most investigations in pure gravity this number
appears to be equal to three \cite{Falls2020}. In deriving such results, one posits that the continuum formalism based
on the metric tensor remains valid all the way to the Planck scale and beyond, at least in some effective sense, and that quantities
like $\Gamma_k$ exist and are well defined.
One challenge in this approach is to identify observables, which is in part related to the issue of background independence,
whose correct implementation remains the main obstacle to progress, both technically and conceptually \cite{Bonanno2020}.
Note that for general values of $k$ neither the couplings $c_{\!\alpha\! } (k)$ nor the average action $\Gamma_k$ are observables, 
or have a direct physical interpretation \cite{Donoghue2019}. The quantity carrying physical information on the
quantum-corrected correlation functions is $\Gamma_{k=0}$, where all quantum fluctuations have been integrated out.
Identifying $k\! \not=\! 0$ with the momentum scale of a specific physical 
application is only meaningful in particular circumstances \cite{Knorr2019}.

To summarize, asymptotic safety is a concrete scenario of how quantum gravity can be well-defined and finite
at arbitrarily high energy, in a setting that is as close to standard quantum field theory as possible. The functional renormalization group methods
used in the asymptotic safety approach provide a versatile computational framework that lends itself for a comparison with 
gravity as an effective field theory \cite{Burgess2003,Donoghue2012,Donoghue2019}. More interesting from a physics
point of view is the formulation's relation with fully nonperturbative approaches, and whether it can accommodate
a nontrivial ground state of quantum geometry, for which there is evidence in CDT quantum gravity (Q27).
This will require a further exploration of the truly nonperturbative aspects of this formulation 
and a further development of effective numerical methods.

\section*{Q24: Why is the gravitational path integral ill-defined?}
\label{sq24}
\addcontentsline{toc}{section}{Q24: Why is the gravitational path integral ill-defined?}

A key quantity in most covariant approaches to quantum gravity is the gravitational path integral
\begin{equation}
Z=\int\limits_{{\cal G}(M)}\!\! {\cal D} [g]\, {\rm e}^{i S^{\rm EH}[g]},
\label{pathintagain}
\end{equation}
also known as the {\it sum over
geometries} or {\it sum over histories}, where for ease of reference we are restating the formal expression for $Z$ from Q7. 
Given the nonrenormalizability of the perturbative path integral (cf.\ Q9), the question is how to make sense of the right-hand side of 
expression (\ref{pathintagain}) for full, four-dimensional quantum gravity beyond perturbation theory.
To understand the magnitude of the challenge, let us review the ingredients that must be specified and the main problems that 
must be overcome to convert
eq.\ (\ref{pathintagain}) from a statement of intent to an actual prescription for computing the gravitational path integral. 
\begin{description}[itemsep=2pt,leftmargin=0pt]
\item[$\bullet$] 
The Lagrangian density that appears in the Einstein-Hilbert action $S^{\text{EH}}$ of eq.\ (\ref{SEH}) is not
quadratic in the metric fields, but contains inverse metrics and a square root of the determinant of the metric, which implies that
the functional integral is not a Gaussian. How then should we compute it?
\item[$\bullet$] Since (\ref{pathintagain}) is an infinite-dimensional functional integral, we expect that it needs to be regularized and
renormalized. This should be done while maintaining the theory's diffeomorphism invariance.  
\item[$\bullet$] The carrier space of the path integral is presumably some distributional version of the quotient 
$\mathcal{G}(M)$.\footnote{For example, in quantum gravity theories based on dynamical triangulations (see Q26 and Q27),
typical histories contributing to the path integral in the continuum limit 
seem to be higher-dimensional analogues of the nowhere differentiable
paths that constitute the carrier space of the path integral of the free nonrelativistic particle, the so-called Wiener measure \cite{Chaichian2001}.}
However, even classically this is not a vector space and it is unclear how to parameterize it. If a gauge-fixing it used, it must be
shown to be unique and attainable, without Gribov ambiguities \cite{Esposito2004}.
\item[$\bullet$] Which diffeomorphism-invariant measure ${\cal D} [g]$ should be used on the space of geometries? Is it unique?
(Note that even if the carrier space had a linear structure, there is no analogue of the Lebesgue measure in infinite dimensions.)   
\item[$\bullet$] The complex nature of (\ref{pathintagain}) makes it not well suited for applying Monte Carlo methods, 
a potentially powerful tool for evaluating the path integral itself and the expectation values of observables.
In quantum field theory on Minkowski space one can invoke the 
Osterwalder-Schrader theorem \cite{Osterwalder1975,Seiler1982} to convert the complex path integral to a real partition function. However, no Wick rotation is known in continuum gravity for general metrics \cite{Dasgupta2001,Visser2017}. 
If one starts from Euclidean quantum gravity, which a priori is a different theory, and tries to compute the 
Euclidean path integral (\ref{eupathint}), one still has to find a way to Wick-rotate any results back to physical, Lorentzian signature.
\item[$\bullet$] The Einstein-Hilbert action in Euclidean signature is unbounded below. This can be illustrated by 
locally adopting proper-time gauge to bring the spacetime metric $g_{\mu\nu}$ to block-diagonal form (such that $g_{00}=\pm 1$, 
$g_{0i}=0$), and subsequently decomposing the spatial metric into a constant-curvature metric $\bar{g}$ and a conformal factor
$\lambda$ according to $g_{ij}(t,\vec{x})=\exp (2\lambda(t,\vec{x}))\bar{g}_{ij}(t,\vec{x})$ \cite{Dasgupta2001}. Substituting this
ansatz into the four-dimensional Einstein-Hilbert action yields
\begin{equation}
S=\frac{1}{16\pi G_{\text N}}\int_M d^4 x\, \sqrt{\det \bar{g}_{ij}}\, {\rm e}^{3\lambda}\left( -6 (\partial_t \lambda)^2+\dots \right),
\label{conformal}
\end{equation}
which demonstrates that the kinetic term of the local conformal mode $\lambda(t,\vec{x})$ enters the action with the {\it wrong}
sign, independent of whether the four-dimensional metric is Lorentzian or Riemannian. This leads to the infamous conformal-mode
divergence, which potentially makes any Euclidean path integral ill-defined, since its weight factors ${\rm e}^{-S^{\rm EH}[g_{\text{eu}}]}$ 
grow without bound \cite{Gibbons1978,Mazur1989}. This is in particular true for Euclidean cosmological path integrals, where the spacetime
metric is truncated to the scale factor $a(t)\!\propto\! \exp \lambda (t)$, which can be identified with the global ($\vec x$-independent)
conformal mode. In the simplest minisuperspace models, this can be dealt with by choosing a suitable integration contour in the complex
$a$-plane. However, it seems difficult to generalize this to a physically compelling and unique prescription
in more complicated cosmological models \cite{Halliwell1989}, let alone in full quantum gravity, since
the conformal factor cannot be isolated globally, i.e.\ in the space of all metrics. 
\end{description}
Any proposal for computing a nonperturbative gravitational path integral has to address these nontrivial technical 
issues or otherwise explain why they are not relevant.

\section*{Q25: What is lattice quantum gravity?}
\label{sq25}
\addcontentsline{toc}{section}{Q25: What is lattice quantum gravity?}

The motivation behind lattice gravity is to emulate the successes of lattice QCD in understanding the nonperturbative sector of 
the theory. In essence, the strategy is to ``put the theory on the lattice'', where the lattice constitutes an approximation 
or regularization of
smooth spacetime, and the {\it lattice spacing} (length $a$ of a lattice edge) provides an ultraviolet (UV) cutoff.
The full quantum theory is
then {\it defined} in the continuum limit $a\!\rightarrow\! 0$, that is, as one approaches a second- (or higher-) order phase transition in the phase
diagram of the lattice theory\footnote{spanned by the bare coupling constants of the lattice theory}, while renormalizing
the bare coupling constants appropriately \cite{Montvay1994,Smit2002}. In this limit, the correlation length of the lattice
theory diverges and the continuum theory exhibits universality, an independence of the details of the (highly nonunique) regularization
that went into setting up the lattice formulation \cite{Goldenfeld1992}.

The crucial question in quantum gravity is whether such a strategy can be meaningfully implemented, given that spacetime geometry itself is dynamical, and whether such continuum limits exist and are reasonably unique. Early attempts made use of a gauge-theoretic, first-order
formulation of general relativity with action
\begin{equation}
S[\omega,e]=\int_M e\wedge e\wedge R[\omega],
\label{gtlagrange}
\end{equation}
with independent $so(3,1)$-valued spin connections $\omega_\mu^{IJ}$ (with curvature $R$) and vierbeins (tetrads) $e_\mu^I$,
or generalizations thereof \cite{Smolin1978,Das1978,Tomboulis1983} (see \cite{Loll1998} for a comprehensive review). 
The important ingredient here is a connection one-form, from which one constructs holonomy lattice variables \`a la Wilson \cite{Wilson1974}. 
Compared to lattice gauge theories, there are several complications when putting such a theory on a lattice, including the unboundedness
of the action.
To yield finite integrations, the connection is ad-hoc compactified to $so(4)$, and also the noncompact vierbein directions must 
be ``tamed'' in some way. In addition, the metricity conditions 
\begin{equation}
\partial_{[\mu}e_{\nu ]}^I +\omega_{[\mu \, J}^I e^J_{\nu ]}=0, 
\label{metricity}
\end{equation}
which classically follow from varying (\ref{gtlagrange}) with respect to $e$, and make sure that $\omega [e]$ is the unique
torsion-free connection associated with $e$, need to be imposed by hand. However, the main reason for classifying these
attempts as na\"ive is that they completely ignore the theory's diffeomorphism symmetry. Since the fields of the continuum theory are
attached in a straightforward manner 
to the discrete edges and vertices of a fixed background lattice, it appears that diffeomorphism invariance is badly broken. 
To summarize, the many uncontrolled features of these lattice gravity models make it difficult to assess how much 
they still have to do with gravity.
At any rate, Monte-Carlo simulations of such models did not find any signs of second-order phase transitions \cite{Caracciolo1989}. 

However, there is another, qualitatively very different way of ``putting gravity on the lattice'', due to Regge \cite{Regge1961},
which is geometric rather than gauge-theoretic and has been much more successful. It is based on a classical approximation 
of smooth, curved spacetimes $(M,g_{\mu\nu})$ by much simpler metric spaces, which are piecewise flat.
These spacetimes are perfectly continuous, but curvature is distributed on them in a discrete way.\footnote{These
spacetimes are therefore not lattices at all, but we will for simplicity adhere to the notion of ``lattice gravity''.} 
The key difference with the gauge-theoretic lattice models, spelled out in greater detail in Q26 below, is that 
this description does not require the introduction of coordinates.
Depending on precisely how this idea is used to parameterize the space ${\mathcal G}(M)$ of all geometries in the gravitational
path integral, it removes part or all of the overcounting due to unphysical degrees of freedom. 
The only nonperturbative path integral approach of this type which both takes the Lorentzian nature of spacetime into account 
{\it and} has found evidence of
second-order phase transitions is that of Causal Dynamical Triangulations (CDT), described further in Q27.

\section*{Q26: What is Regge calculus and why is it important for quantum gravity?}
\label{sq26}
\addcontentsline{toc}{section}{Q26: What is Regge calculus and why is it important for quantum gravity?}

\begin{figure}[t]
\centerline{\scalebox{0.5}{\rotatebox{0}{\includegraphics{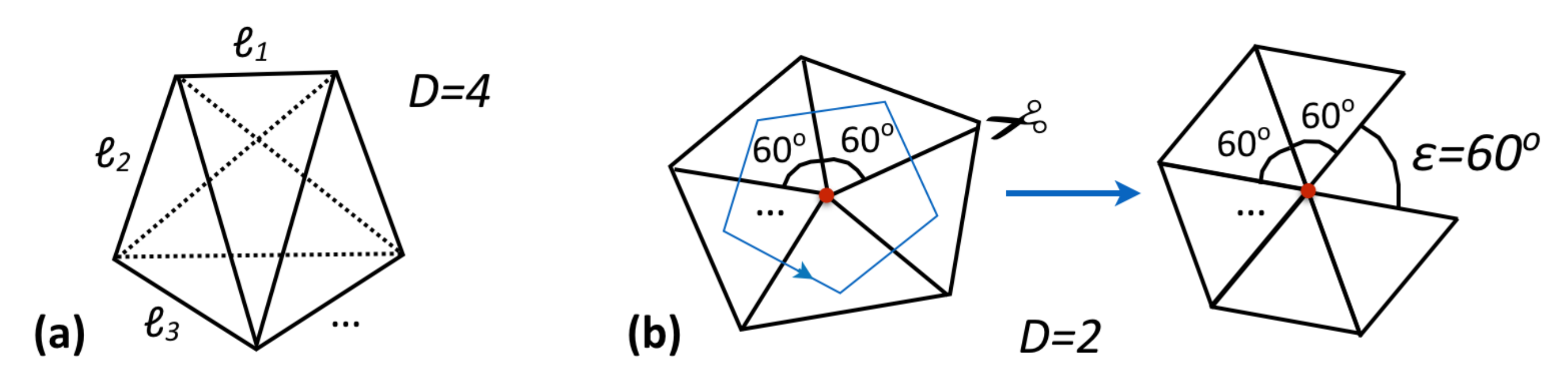}}}}
\caption{(a) A four-simplex, whose ten edge lengths $\ell_i$ determine its interior flat geometry uniquely. (b) Gluing together five
equilateral triangles yields a two-dimensional disc with a curvature singularity (with Gaussian curvature $\pi/3 =60^\degree$) 
at the central vertex. The disc is flat everywhere else.}
\label{reggetriangles}
\end{figure}

The foundations of Regge calculus were laid in 1961 in a paper entitled ``General relativity without coordinates'' \cite{Regge1961} 
(see \cite{Misner1973} for a pedagogical introduction).
It was motivated by the wish to formulate and solve the classical Einstein equations -- especially in situations without a large 
degree of symmetry -- in an approximate way, without having to deal 
with coordinate redundancies. In this formulation, a spacetime is assembled from four-simplices. 
A four-simplex (Fig.\ \ref{reggetriangles}a) 
is a generalization to $D\! =\! 4$ of a triangle ($D\! =\! 2$) and a tetrahedron  ($D\! =\! 3$), whose interior by definition is flat,  
i.e.\ has vanishing Riemann
tensor. Its geometry is determined {\it uniquely} by the values of its ten squared edge lengths $\ell_i^2$, $i\! \in\! [1,10]$, which
can be time-, space- or light-like. A four-simplex can be thought of as a piece of spacetime cut out of Minkowski space. 
Note that the framework of Regge calculus is applicable in any dimension and signature; 
in Riemannian signature, all squared edge lengths of a $D$-simplex are necessarily spacelike, i.e.\ positive. 

Since the individual simplicial building blocks are flat, curvature can only arise when assembling them into a larger simplicial manifold, 
by gluing
together $D$-simplices pairwise along matching $(D\! -\! 1)$-dimensional flat simplicial faces, which in $D\! =\! 4$ are tetrahedra. 
In this way one obtains piecewise flat spaces whose curvature is concentrated at sub-simplices of dimension $D\! -\! 2$. This
is most easily illustrated in two dimensions. Consider for simplicity a set of five identical, equilateral Euclidean triangles and glue
them together pairwise along one-dimensional faces (edges) around a central vertex (Fig.\ \ref{reggetriangles}b). The result
is a two-dimensional simplicial manifold with the topology of a disc, whose central vertex (a 0-simplex) carries a Gaussian curvature of $+\pi/3$. 
Importantly, this curvature is intrinsic to the surface (independent of any embedding) and can be determined by parallel-transporting 
a tangent vector along any closed curve enclosing the vertex. The presence of curvature can also be visualized by referring 
to an explicit embedding into
Euclidean space: cut open the disc along one of its internal edges and put it onto a flat two-dimensional 
surface (Fig.\ \ref{reggetriangles}b) to exhibit a {\it deficit angle} $\varepsilon$, which is the angle ``missing to $2\pi$'' and
again a direct measure of the intrinsic, Gaussian curvature located at the vertex. In four dimensions, curvature is concentrated 
at two-dimensional triangles, and captures the Gaussian curvature of a two-surface perpendicular to such a triangle. 

The metric information, which in the continuum is contained in the smooth field tensor $g_{\mu\nu}(x)$, 
in Regge calculus is contained in the edge
length assignments $\{ \ell^2_i\}$ and the gluing data (connectivity matrix) describing the underlying simplicial manifold $T$.
Regge calculus has been used to a limited extent in classical $(3\! +\! 1)$-dimensional evolution schemes in numerical relativity (see \cite{Williams1991,Gentle2002,Barrett2018} and references therein), but overall does not seem to yield advantages over other methods. 
This situation changes dramatically in the nonperturbative, covariant quantum theory. The reason is the
manifest diffeomorphism invariance of this formulation, which resolves one of the key problems in dealing with the path 
integral (cf.\ Q24), a fact that generally has been underappreciated by the community.

The first approach of this kind was quantum Regge calculus \cite{Williams1991,Hamber2009,Frohlich1981}, whose
starting point is the Regge action $S^{\text{R}}(T,\{ \ell_i^2\})$, a simplicial version of the 
Einstein-Hilbert action \cite{Regge1961}. One implements a regularized version of the path integral by fixing a 
finite triangulation $T$ (whose size should eventually be taken to infinity) and integrating over all edge lengths, in the presence of
some length cutoff. The fact that the $\ell_i^2$ have to satisfy triangle inequalities 
makes an analytic evaluation of the nonperturbative path integral impossible, in any dimension. However, it does not impede 
numerical simulations of the path integral of Euclidean quantum Regge calculus, which were actively pursued between the
mid-1980s \cite{Berg1985,Hamber1985} and the late 1990s \cite{Hamber1999}. They broadly agree on the presence of a phase of 
smooth geometry and one of rough geometry, with a range of findings for other properties of the model (see \cite{Loll1998} for an overview).
The most recent numerical results on the status of the phase transition and the extraction of certain critical exponents \cite{Hamber1999} 
are still awaiting independent confirmation. 

A complementary way of performing the gravitational path integral is that of Dynamical Triangulations (DT), where one fixes all edge lengths to
the same value\footnote{This is true for Euclidean DT. In Lorentzian, causal DT one allows for two different values, to be able to distinguish between space- and time-like edges.} $\ell_i\!\equiv\! a$, which at the same time serves as an ultraviolet
cutoff. The sum over geometries is implemented by summing over all triangulations $T$, i.e.\ all inequivalent ways of gluing
together identical equilateral four-simplices to obtain a simplicial manifold of a given topology \cite{Ambjorn1992}. DT quantum gravity 
improves in several ways on quantum Regge calculus: there is no residual gauge invariance \cite{Rocek1982},   
the unboundedness of the action is cured nonperturbatively \cite{Dasgupta2001,Ambjorn2002}, and there is a
Lorentzian version with a well-defined Wick rotation that is defined on the piecewise flat geometries, namely, 
CDT quantum gravity \cite{Ambjorn2013,Loll2019}. As already mentioned in Q17,  
DT quantum gravity models in $D\! =\! 2$ can be solved analytically for either signature and are equivalent to known continuum descriptions 
(which of course are not available in $D\! =\! 4$), boosting confidence in the approach as such.   

First results on the numerical evaluation of the four-dimensional Euclidean DT path integral, 
\begin{equation}
Z^{\text{DT}}=\lim_{a\rightarrow 0}\; \sum\limits_{\text{Euclidean}\atop \text{triang.}\, T}\frac{1}{C(T)}\, {\rm e}^{-S^{\text R}[T]},
\label{dtpi}
\end{equation}
appeared 
in 1991 \cite{Agishtein1991,Ambjorn1991}, and were followed by a boost of 
activity (see \cite{Loll1998} for an overview of results).\footnote{The measure in the path integral (\ref{dtpi}) is 
``democratic'', in the sense that each triangulation $T$ carries the same weight, up to the factor $C(T)$, denoting
the size of the automorphism group of $T$, a natural number almost always equal to 1.}
They found evidence of a two-phase structure, consisting of a crumpled and a branched-polymer phase, with a third, ''crinkled'' phase appearing
when a nontrivial measure term is included in the path integral \cite{Bilke1998}. However, neither during the 1990s nor in 
recent re-examinations of the Euclidean version of DT quantum gravity evidence of a second-order phase transition has been 
found \cite{Ambjorn2013a,Coumbe2014,Rindlisbacher2015}, which makes it difficult to attach a standard continuum interpretation to this
model, apart from the challenges of interpreting it in Lorentzian terms or establishing its unitarity, given the absence of reflection positivity.

\section*{Q27: What are causal dynamical triangulations?}
\label{sq27}
\addcontentsline{toc}{section}{Q27: What are causal dynamical triangulations?}

Causal dynamical triangulations (CDT) were introduced as an attempt to make nonperturbative quantum gravity more
physical (i.e.\ to allow for notions of time and causal structure, unlike in Euclidean models) {\it and} come up with quantitative
evidence for nontrivial continuum limits \cite{Ambjorn1998,Ambjorn2000}.
One can simply think of CDT as a ``Lorentzian variant'' of Euclidean DT, but this understates the challenge one usually
faces in quantum gravity, namely, to describe a quantum theory of {\it Lorentzian} spacetimes while maintaining access to powerful computational tools that are needed to extract the physical properties of these theories, which typically requires a
{\it real, Euclidean} version of the path integral. In CDT, both aspects are realized due to the presence of
a well-defined ``Wick rotation'' \cite{Ambjorn2001,Loll2019}, an analytic continuation in one of the geometric parameters of the model, 
which maps the regularized Lorentzian path integral
\begin{equation}
Z^{\text{CDT}}=\lim_{a\rightarrow 0}\; \sum\limits_{\text{Lorentzian}\atop \text{triang.}\, T}\frac{1}{C(T)}\, {\rm e}^{iS^{\text R}[T]}
\label{cdtpi}
\end{equation}
to a real partition function which {\it can} be investigated with numerical approximation methods. Note that this
partition function is different from the Euclidean path integral (\ref{dtpi}), because the sum is over a different ensemble of triangulations, 
which -- importantly -- leads to inequivalent outcomes. 

The CDT path integral addresses all difficulties raised in Q24. All elements of its
construction are defined on a regularized version of the space ${\mathcal G}(M)$ of geometries, and the path integral has no
residual gauge invariance. 
Unless the bare inverse Newton constant is chosen too large, the conformal divergence (\ref{conformal}) coming from the 
gravitational action is entropically suppressed: configurations that maximize the kinetic term exist, but are relatively rare in the nonperturbative ensemble in the continuum limit \cite{Ambjorn2010a,Dasgupta2001}. 
The regularized geometries carry a natural discrete time parameter, with respect to which reflection
positivity has been shown \cite{Ambjorn2001}, and which on sufficiently large scales turns out to be proportional to 
cosmological proper time \cite{Ambjorn2007,Ambjorn2008}. The presence of an explicit time parameter is the simplest 
way to enforce global hyperbolicity, which is an important feature that distinguishes the CDT from the DT ensemble.
However, explicit studies in $D\! =\! 3$ suggest that a time foliation is not strictly necessary to achieve this \cite{Jordan2013,Jordan2013a}.\footnote{Note that, 
unlike in the continuum, e.g.\ in the $(3\! +\! 1)$-decomposition of canonical gravity (cf.\ Q19) or in 
Ho\v rava-Lifshitz gravity \cite{Horava2009}, a choice of time here
is not related to any breaking of diffeomorphism invariance, since this invariance is already manifest. Also, it is not a priori
associated with any physical notion of time, see \cite{Loll2019} for further discussion.}

Like DT, also CDT quantum gravity is amenable to Markov chain Monte Carlo methods and has already been studied extensively (see
\cite{Ambjorn2012,Loll2019,Ambjorn2021} for reviews). 
The model has a nontrivial phase structure and two lines of second-order phase transitions \cite{Ambjorn2011,Ambjorn2012a,Coumbe2015,Ambjorn2016,Ambjorn2017}, which provide candidates for 
critical points where a continuum theory may be defined. For topology\footnote{For technical convenience, time 
in the computer simulations is often compactified.} $M\! =\! S^1\!\times\! S^3$, at least one of the observed phases, the so-called  
de Sitter phase, features an emergent four-dimensional geometry \cite{Ambjorn2004,Ambjorn2005a}, which is a highly nontrivial result 
as will be explained in Q28. 
In the sense of expectation values, the large-scale shape of the emergent geometry is that of a classical de Sitter space
\cite{Ambjorn2007,Ambjorn2008,Ambjorn2011a}. Likewise, its large-scale curvature properties are compatible with those of
a de Sitter space \cite{Klitgaard2020}, despite being extremely quantum-fluctuating on short scales. 

There is a proof-of-concept study of renormalization group flows in the CDT phase space \cite{Ambjorn2014}, but 
mainly for technical reasons it has not yet been possible to make definite statements about these flows
near the second-order phase transitions. This is needed to better understand the nature of the UV completion and
whether or not the asymptotic safety scenario (cf.\ Q23) can be corroborated. 
Irrespectively, CDT appears to be a valid calculational framework that makes it possible to measure
quantum observables within an order of magnitude of the Planck scale (see Q28, Q29), uncovering ``Planckian fingerprints'' 
\cite{Ambjorn2005,Ambjorn2008} as well
as extracting classical limits \cite{Ambjorn2004a}. This constitutes considerable progress in a field where traditionally there were no operationally
well-defined, approach-independent quantities to be computed. One central task of nonperturbative quantum gravity is to
uncover more of such diffeomorphism-invariant observables. Lastly, it should be noted that (C)DT quantum gravity theories
have very few free parameters and a large degree of universality. This means they can potentially be falsified, a highly 
desirable property for any candidate quantum gravity theory to have. For this, it is not necessarily required 
to develop testable phenomenology that probes the deep quantum regime

\section*{Q28: How can the dimension of quantum spacetime \textit{\textbf{not}} be 4?}
\label{sq28}
\addcontentsline{toc}{section}{Q28: How can the dimension of quantum spacetime \textit{\textbf{not}} be 4?}

One of the most challenging and exciting tasks in nonperturbative quantum gravity is to understand
the nature of quantum geometry near the Planck scale and to develop appropriate notions (read ``quantum observables'')
and approximation tools to explore this terra incognita. Here, the word ``geometry'' should be understood in a generalized
sense, since we do not know a priori which attributes of classical spacetime geometry, if any, will continue to be
meaningful in a Planckian regime. 

Fortunately, as already mentioned in Q14, many important lessons have been learned about
quantum-geometric systems over the last decades,
primarily in the context of random geometry, which includes models of (causal) dynamical triangulations in 
various dimensions,
but also studies of causal set theory \cite{Surya2019} and tensor models \cite{Rivasseau2013,Gurau2016}. 
For systems of random geometry the logic is straightforward: 
elementary building blocks of geometry are assembled into metric spaces, which are then
made to interact according to the rules of quantum field theory, i.e.\ superposed in a path integral, where each geometry 
appears with an amplitude weight that depends on its gravitational action, as in (\ref{dtpi}) and (\ref{cdtpi}).

The reason why such simple, non-exotic ingredients can lead to dynamical systems with nontrivial and sometimes
very unexpected properties has to do with their nonperturbative collective behaviour, which at the level of their partition functions 
is governed by a complex interplay between energy and entropy,  
and with the fact that their construction involves a nontrivial limiting process.
In fact, one of the unexpected and unwelcome lessons one has learned is that in $D\! \geq \! 3$, systems of Riemannian random
geometries, made from $D$-dimensional building blocks, generically exhibit geometrically degenerate behaviour that does not
allow for an interpretation in terms of extended, $D$-dimensional geometry on any scale above the cutoff scale. 

How is it possible that a superposition of spacetimes, each one assembled from four-dimensional
simplices, generates a quantum spacetime that is {\it not} four-dimensional? One way in which this can happen is
that in a particular
region of phase space the four-simplices preferably arrange themselves into a
structure that at the cutoff scale $a$ is a four-dimensional manifold (by construction), but when viewed from a
larger scale\footnote{any finite {\it physical} scale, after making the lattice infinitely fine, i.e.\ in the limit $a\rightarrow 0$} 
looks like a {\it branched polymer}. 
The latter is a well-known statistical system, characterized by a Hausdorff dimension of 2
and a spectral dimension of 4/3 on all scales.
In other words, the dimension of the microscopic building blocks does not determine the dimension of the emergent
``quantum spacetime''!

Unfortunately, the properties of branched polymers are incompatible with the existence of a classical limit that reproduces the
extended spacetimes of general relativity.
Over time, it has been understood that the occurrence of this pathological behaviour is
generic, and is not only present in DT quantum gravity, where it does not even depend on the dimensionality of the simplices \cite{Ambjorn1995,Ambjorn1991a,Veselov2003},
but also beyond \cite{Gurau2013}. Polymerization is not the only type of degeneracy afflicting ensembles of
Euclidean random geometry beyond $D\! =\! 2$ and potentially standing in the way of a large-scale gravity interpretation \cite{Hotta1995,Catterall1995}. However, requiring a well-defined {\it causal structure} to be present for path integral histories, as is done
in CDT quantum gravity, gives rise to a phase structure which is not entirely dominated by such degeneracies, but
allows for the presence of behaviour compatible with four-dimensionality on large scales, as we will describe below.

The above discussion highlights the important role of {\it dimensionalities} as diffeomorphism-invariant observables characterizing 
spacetimes beyond classicality. While superstring theory has taught us not to take spacetime dimension 4 for granted microscopically, 
the dynamical mechanism at work here is quite different and directly related to the presence 
of large quantum fluctuations on short scales and the absence of any a priori preferred classical background.
As is illustrated by the example of the branched-polymer behaviour, in nonperturbative quantum gravity 
these dimensions will generically not assume their canonical, classical value of 4, and may not even be an integer, which
implies that they are not associated with {\it any} classical space.
The latter is perfectly acceptable in a deep quantum regime, but not in the limit of large, classical distances. 

Another nonperturbative and qualitatively new phenomenon is that the dimension of spacetime can be {\it scale-dependent}. 
This was first discovered in CDT quantum gravity \cite{Ambjorn2005} for the spectral dimension $D_S$, which associates a dimension 
with the behaviour of a diffusion process. On a compact manifold $M$ of volume $V(M)$, one considers the 
probability $P(x,x,\sigma)$ of a random walker to return to a point $x\!\in\! M$ after a diffusion time $\sigma$.
To obtain a diffeomorphism-invariant quantity, one integrates $P$ over the manifold to obtain the average return probability
\begin{equation}
\mathcal{R}_V (\sigma)=\frac{1}{V(M)}\int \!\! d^4x\ \sqrt{\det g} \, P(x,x,\sigma)\sim\frac{1}{\sigma^{D_S/2}},
\label{return}
\end{equation}
from which the spectral dimension $D_S$ is extracted as (minus two times) the power 
of the leading term in a small-$\sigma$ expansion \cite{Ambjorn2012}. 
When evaluating (\ref{return}) on the ensemble of CDT configurations in $D\! =\! 4$, one finds that the expectation value
$\langle D_S \rangle$ on large scales is given by $4.02\pm 0.1$, i.e.\ compatible with 4, signalling a correct classical limit, 
but near the Planck scale 
drops smoothly to $1.80 \pm 0.25$, which is compatible with the value 2 within measurement 
accuracy \cite{Ambjorn2005,Ambjorn2005a}. The scale
in this context is set by $\sigma$, because the typical maximal linear distance away from the point $x$ reached by the random 
walker is of the order of $\sqrt{\sigma}$. 

Following the discovery of this {\it dynamical dimensional reduction} near the Planck scale, many arguments and 
computations have corroborated this phenomenon from the perspective of other quantum gravity approaches
(see e.g.\ \cite{Modesto2008,Horava2009a,Reuter2011,Rechenberger2012,Carlip2015}). What makes the short-scale spectral dimension
such an attractive quantity is the fact that it can be computed across a whole range of candidate theories, in a field largely
devoid of any computable ``numbers''. It has been suggested \cite{Carlip2017} that dimensional reduction may
be a universal feature of nonperturbative quantum gravity\footnote{see \cite{Loll2019} for some cautionary remarks}, 
which means that $D_S\! =\! 2$ could have a status similar to
the famous formula $S_{\text{BH}}\! =\! A/4$ for the entropy of a black hole of area $A$, which by contrast is a prediction 
of semi-classical gravity. 
Despite the fact that it is currently difficult to derive phenomenological consequences
of a Planck-scale dimensional reduction in a stringent manner,
the spectral dimension is already being used as a benchmark for comparing different
quantum theories of gravity.

\section*{Q29: What (other) quantum observables are there?}
\label{sq29}
\addcontentsline{toc}{section}{Q29: What (other) quantum observables are there?}

When introducing observables in pure gravity in Q6, we explained that they are necessarily nonlocal
in nature, as a consequence of background independence. The same holds at the level of the quantum theory. 
For example, as we saw in eq.\ (\ref{return}), the extraction of the spectral dimension involves a spacetime average.
For this or any other observable ${\mathcal O}[g]$, its expectation value $\langle {\mathcal O} \rangle$ 
in the nonperturbative quantum theory is obtained by taking an additional ensemble average,
\begin{equation}
\langle {\mathcal O} \rangle =\frac{1}{Z} \int\limits_{{\cal G}(M)}\!\! {\cal D} [g]\, {\mathcal O}[g]\, {\rm e}^{i S^{\rm EH}[g]},
\label{eigenvalue}
\end{equation}
where for simplicity we are using a formal continuum language, and $Z$ was defined in eq.\ (\ref{pathintagain}) above.
Although it may appear natural to use observables ${\mathcal O}[g]$ which are spacetime integrals of scalars constructed from the
Riemann tensor and its covariant derivatives, this is in general not feasible, both technically and as a matter of principle.
In a continuum formulation, each instance of the Riemann tensor contains double derivatives of the metric at the same point, which
must be regularized and renormalized while respecting diffeomorphism invariance, a highly nontrivial requirement. 
Taming of the associated infinities may not be possible, and may introduce ambiguities or an uncontrolled gauge dependence. 
In discrete or piecewise flat nonperturbative formulations, tensor calculus is generally not available and derivatives are not defined. 

Due to these difficulties, diffeomorphism-invariant $n$-point functions\footnote{where the locations of the $n$ points are integrated over, 
subject to fixed pairwise geodesic distances} of curvature scalars are for the most part beyond reach in four dimensions, although they would in 
theory provide for a rich set of observables. 
The quantum observables that have been constructed and studied so far rely on notions of geodesic distance
and volumes, which do not require the presence of a smooth geometric structure, and have a relatively simple functional
form in terms of these basic quantities. Of course, the need for a proper
renormalization of physical observables persists.
Other considerations regarding feasibility have to do with computational and numerical limitations. 
For models of lattice gravity, these relate e.g. to system size, computing power and the efficiency of the Monte Carlo algorithms.

Given these multiple requirements, it is rather nontrivial that a number of quantum observables have nevertheless
been identified and investigated:
\begin{description}[itemsep=2pt,leftmargin=0pt]
\item[$\bullet$] 
A notion of dimension different from the spectral dimension of Q28 is the (local) {\it Hausdorff dimension} $d_H$, which was
already mentioned in Q17.
It is determined from the scaling behaviour of the volume of a geodesic ball ${\mathcal B}_R(x)$  
of geodesic radius $R$ centred at $x\!\in\! M$. 
To obtain a diffeomorphism-invariant quantity, the ball volume $\text{vol}({\mathcal B}_R(x))$ is averaged over $M$, 
and $d_H$ is defined by the leading power of its expectation value 
\begin{equation}
\langle\,  \overline{\text{vol}({\mathcal B}_R)}\, \rangle  \sim R^{d_H}
\label{haus}
\end{equation}
for small $R$. A large-scale or {\it cosmological Hausdorff dimension} can be extracted by relating the average linear size 
of a spacetime to its total volume, for different volumes. In the de Sitter phase of CDT quantum gravity it was
determined to be 4 within measuring accuracy \cite{Ambjorn2005a,Ambjorn2012}, signalling a good classical limit.
\item[$\bullet$]     
A geometric observable of a global nature is the {\it volume profile} or {\it shape} $V_3(\tau)$ of the spacetime, which is given
by the three-volume of spatial slices as a function of (cosmological) proper time $\tau$. The latter is a geometric concept,
obtained by measuring the geodesic distance from an initial surface or an initial singularity,
emulating a corresponding classical construction \cite{Andersson1997}. In CDT quantum gravity with 
spacetime topology $M\! =\! S^1\!\times\! S^3$, the expectation value $\langle V_3(\tau)\rangle $ can
be mapped onto the volume profile of a (Euclidean) de Sitter space with great accuracy \cite{Ambjorn2007,Ambjorn2008,Ambjorn2011a},
while for spacetime topology $M\! =\! S^1\!\times\! T^3$ the volume profile is constant in $\tau$ \cite{Ambjorn2016a}.
Remarkably, both nonperturbative models therefore reproduce the scale factor of a solution of a classical cosmological 
minisuperspace model
in the sense of expectation values \cite{Loll2019,Glaser2017}.\footnote{Note that the three-volume satisfies 
$V_3(\tau)\!\propto\! a^3(\tau)$, where $a(\tau)$ is the scale factor of cosmology mentioned in Q17.} 
Moreover, the dynamics of the {\it quantum fluctuations} $\delta V_3(\tau )$ of the three-volume around its mean 
measured in simulations is consistent with a semiclassical analysis of the corresponding quantum fluctuations
in the continuum \cite{Ambjorn2008,Ambjorn2012}. The quantum effective action for the scale factor can also be
studied in asymptotic safety, with consistent findings \cite{Knorr2018}. 
\item[$\bullet$]     
The most recent ingredient in constructing quantum observables is a notion of Ricci curvature applicable
to nonsmooth metric spaces. This quantum Ricci curvature, inspired by
the Ollivier-Ricci curvature in pure mathematics \cite{Ollivier2009}, is extracted by comparing the distance between
two geodesic spheres with the distance of their respective centres \cite{Klitgaard2017,Klitgaard2018}. 
The simplest observable one can construct from it is 
the so-called {\it curvature profile} $\bar{K}_\delta$ \cite{Brunekreef2020}, the spacetime average of a quantity $K_\delta (x)$, 
which has the interpretation of a Ricci scalar coarse-grained over a
neighbourhood of diameter $\delta$ at the point $x$. Evaluating its expectation value $\langle\, \bar{K}_\delta\,\rangle$
gives rise to a scale-dependent, renormalized Ricci curvature that turns out to be well defined in a Planckian regime. It has been used
in CDT quantum gravity to show that the averaged scalar curvature of its emergent quantum geometry 
is compatible with that of a classical de Sitter space. This further strengthens the evidence for the existence of a classical limit of
this particular quantum gravity theory, and paves the way for more elaborate curvature
observables, which also use the directional character of the quantum Ricci curvature \cite{Klitgaard2020}.
\end{description}

These examples illustrate the nature and scope of the type of quantum observable one has access to in nonperturbative quantum gravity.
The fact that these quantities are very different from those one typically studies in classical general relativity is not a bug, but a
feature of Planckian physics. This is a regime where most questions one asks classically are simply not operationally well
defined, because of the absence of a classical background and any preferred coordinate system, and the presence of large
quantum fluctuations. 

The entropy of a black hole is a case in point: to make sense of this notion in
the nonperturbative quantum theory would in the first instance require us to develop a diffeomorphism-invariant
characterization of what constitutes a black hole in this context. This is not at all straightforward, since the usual treatments
use a semiclassical setting of quantum fields on a (quasi-)classical background geometry and do not involve the full, nonperturbative
dynamics of the gravitational field itself. By the same token, there are no obvious, renormalized Planckian analogues 
of event horizons\footnote{see 
\cite{Dittrich2005} for exploratory considerations of how to construct a ``quantum horizon finder'' in CDT}, areas and entropy.
Secondly, the question is whether quantum black holes, in whichever way defined, can be shown to arise dynamically in quantum gravity, 
in a regime we have computational access to. This may depend e.g.\ on the choice of boundary conditions for the path integral, but
at the present stage little is known about this in any nonperturbative approach. For a related perspective on black holes and their
entropy in the asymptotic safety scenario, see \cite{Bonanno2020}.

The ongoing challenge of quantum gravity is to improve our calculational abilities to systematically explore existing observables and 
to identify new ones. When working in lattice approaches, it should be noted that by far not every coordinate- or labeling-invariant quantity 
one can define on finite lattices (e.g.\ by counting lattice substructures of a certain type) will lead to an interesting observable in the continuum limit. Many such quantities will be discretization artefacts, in the sense of not exhibiting interesting scaling behaviour in 
a finite-size scaling analysis \cite{Goldenfeld1992,Newman1999} as the lattice size increases. This also implies that they will not
have a large-scale classical limit. 

Lastly, let us comment on the nonlocal nature of purely gravitational observables. It is sometimes suggested to use relational
observables of the type ``scalar quantity $X$ at the point where scalar quantity $Y$ assumes the value 5.8''. This localizes the quantity
$X$ in the point(s) where $Y\! =\! 5.8$, but does not make the corresponding quantum observable local in an operational sense,
since each configuration in the sum over geometries has to be scanned globally to find these points. In addition, such
observables tend to be functionally involved, which also hinders their implementation in a Planckian regime. A different way forward
to close the gap between (semi-) classical and nonperturbative quantum gravity is to compute the nonlocal, integrated observables
that are used in the quantum theory (like the curvature and volume profiles mentioned above) on selected {\it classical} curved
manifolds. This
seems particularly suggestive for the early universe, which is still (relatively) small, and where one would like to compare 
nonperturbative findings with continuum results obtained on a fixed cosmological background.

\section*{Q30: What is the future of quantum gravity?}
\label{sq30}
\addcontentsline{toc}{section}{Q30: What is the future of quantum gravity?}

An overall picture emerges from the panorama of questions and answers we have presented in these lecture notes. 
After 35 years of excursions into the world of extended fundamental objects, quantum gravity has entered a
post-loop and post-string era. 
We are witnessing a renaissance of quantum field theory\footnote{see \cite{Draper2020} for a recent example in a post-string spirit}, 
with the benefit of
computational techniques that have been developed during roughly the same period, to deal with the dynamical nature of gravity
and to explore the nonperturbative sector of the quantum theory. Covariant, path-integral methods are being employed almost 
universally. 

In pursuit of constructing a nonperturbative quantum gravity theory valid on all scales, without relying 
on the presence of asymptotically flat or stationary regions of spacetime, we examined which types of tools can get us there.
Given the state of the art in understanding interacting quantum field theories in four dimensions beyond perturbation theory, 
analytical methods alone are highly unlikely to provide the answer. 
In the absence of guidance from experiment, numerical tools are therefore essential to develop and test
our candidate theories of quantum gravity quantitatively.

Progress in quantum gravity will depend on the continued development of nonperturbative computational methods, hand in hand
with theoretical and analytical considerations.
Approaches that lend themselves to a numerical treatment have a clear advantage over others. 
We introduced the most promising lattice gravity approaches currently on the market, 
which are based on the powerful idea of working directly with spaces of geometries, avoiding the use of coordinates.
The aim of both lattice formulations and continuum treatments in the context of asymptotic safety is to investigate how a
nonperturbative renormalization \`a la Wilson can be realized in quantum gravity \cite{Ambjorn2020,Bonanno2020}, 
ideally complementing each other. Concrete, related research goals are to further strengthen
the evidence for an ultraviolet fixed point, and to explore 
the universal and nonuniversal features of physics along renormalization group trajectories running into the fixed point, using suitable observables. 
Diffeomorphism-invariant observables are absolutely essential to progress; however, we must keep in mind
that quantities which can be meaningfully studied in a Planckian regime tend to differ from their classical
counterparts (cf.\ Q29). 

Tackling these issues is not at all trivial, but the fact that there is a blueprint for how to proceed, which is not tied to any
single approach, is a sign of the growing maturity of the field. 
For example, any candidate theory of quantum gravity with a sufficient degree of uniqueness,
and whose computational capabilities beyond perturbation theory are sufficiently developed 
may provide further evidence for or against the currently favoured high-energy scenario of asymptotic safety. 

We have also highlighted why quantum gravity and the structure of its nonperturbative ground state 
at the Planck scale can be very nontrivial, despite the absence of any exotic ingredients and the adherence 
to the principles of quantum field theory, albeit in a background-independent context.
This is illustrated by two-dimensional toy models of quantum (random) geometry, which have a
rigorous mathematical underpinning, and exhibit highly nonclassical behaviour (cf.\ Q17, Q18). In full, four-dimensional 
quantum gravity, the scale-dependence of the spectral dimension (see Q28) has given us a first glimpse
of the rich world of dynamical quantum geometry, where many other nonperturbative surprises may be awaiting us. 
Not long ago we had little idea that these structures existed {\it and} at the same time could have a well-defined
classical limit, since we lacked some theoretical insights\footnote{We are referring to the introduction of a causal
structure \`a la CDT in lattice gravity. In general, the role of the metric signature and the status of unitarity
in nonperturbative quantum gravity need to be understood better \cite{Loll2019,Donoghue2019,Bonanno2020}.}
and the appropriate computational tools to explore either aspect.

This situation has improved significantly: 
although by far not a complete set, we now have a number of quantum observables at our disposal (Q29),
which allow us to understand selected (near-)Planckian properties of quantum gravity in quantitative terms. 
Their analysis provides both potential benchmarks for comparing different approaches and stepping stones for 
further development, most importantly, to determine
to what extent any microscopic findings can lead to physical consequences on more macroscopic scales. 
A particularly fruitful area of research may be the very early universe, where one can look for a dynamical, 
quantum-gravitational origin of ``spacetime as we know it'' and for purely gravitational sources of structure formation.  
These are challenging and ambitious goals, which promise a high reward and -- if nature is kind to us -- may be just within 
the reach of a theory of quantum gravity.

\end{document}